\documentclass[aps,prapplied,superscriptaddress,footinbib,floatfix,twocolumn]{revtex4-2}
\usepackage{graphicx}
\usepackage{multirow}
\usepackage{color}
\usepackage{amsmath}
\usepackage{amsfonts}
\usepackage{amssymb}
\usepackage{graphicx}
\graphicspath{{Figures/}}
\usepackage{epstopdf}
\usepackage{empheq}
\usepackage{float}
\usepackage{cases}
\usepackage{mathtools}
\usepackage{dcolumn}
\usepackage{bbold}
\usepackage{bm}
\usepackage{ulem}
\usepackage[T1]{fontenc}
\usepackage{gensymb}

\usepackage[dvipsnames]{xcolor}
\usepackage[colorlinks]{hyperref}
\hypersetup{
colorlinks=True,
linkbordercolor=White,
linkcolor=BrickRed,
citecolor=Blue,	 
}

\DeclarePairedDelimiter\ket{\lvert}{\rangle}
\DeclarePairedDelimiter\braket{\langle}{\rangle}

\begin{document}

\title{Disorder-independent hole spin manipulation by hopping}

\author{Biel Martinez}
\email{biel.martinezidiaz@cea.fr}
\affiliation{Univ. Grenoble Alpes, CEA, Leti, F-38000, Grenoble, France}

\author{Ana Sempere-Sanchis}
\affiliation{Univ. Grenoble Alpes, CEA, IRIG-MEM-L\_Sim, F-38000, Grenoble, France}

\author{Jos\'e C. Abadillo-Uriel}
\affiliation{Quantum Advanced Research Center (QuARC), Consejo Superior de Investigaciones Cient{\'i}ficas (CSIC), Sor Juana In{\'e}s de la Cruz 3, 28049 Madrid, Spain}
\affiliation{Instituto de Ciencia de Materiales de Madrid (ICMM), Consejo Superior de Investigaciones Científicas (CSIC), Sor Juana Inés de la Cruz 3, 28049 Madrid, Spain}%

\author{Yann-Michel Niquet}
\email{yniquet@cea.fr}
\affiliation{Univ. Grenoble Alpes, CEA, IRIG-MEM-L\_Sim, F-38000, Grenoble, France}

\date{\today}

\begin{abstract}
Spin manipulation by hopping has recently emerged as a promising strategy to control hole spins in quantum dots using exclusively baseband control, thereby mitigating power dissipation and high-frequency management constraints in large-scale architectures. Unlike conventional approaches such as electric- dipole spin resonance (EDSR), this mechanism exploits dot-to-dot variations of the spin precession axes to enable spin rotations. However, it is intrinsically disorder-dependent: in the absence of sufficient variability, the precession axes remain aligned and spin manipulation becomes ineffective. This fundamental reliance on disorder raises concerns regarding its compatibility with the long-term evolution of spin-qubit platforms toward improved material quality, cleaner interfaces, and enhanced device reproducibility. Here, we numerically assess the viability of spin manipulation by hopping as a function of disorder strength and demonstrate that its implementation is indeed increasingly constrained as disorder is reduced. To overcome this limitation, we propose an alternative strategy based on hopping between intentionally squeezed quantum dots. This approach retains the advantages of baseband control while being independent of disorder and robust against moderate variability, thereby offering improved prospects for scalable hole-spin quantum computing architectures.
\end{abstract}

\maketitle

\section{Introduction}

Semiconductor spin qubits have demonstrated sustained and remarkable progress, establishing themselves as one of the most promising platforms for quantum computing \cite{Loss1998,Petta2005,Maurand2016,Hendrickx2018,Hendrickx2020,Hendrickx2021,Philips2022}. High-fidelity one- and two-qubit gates exceeding the error-correction threshold have been systematically achieved using both electron and hole spins confined in quantum dots (QDs) across multiple material platforms \cite{Yoneda2018,Xue2022,Steinacker2024,Huang2024}. At the same time, the number of qubits integrated into quantum processors continues to scale up \cite{Takeda2021,Philips2022,john2024,tosato2025,nickl2025,seidler2025}, while technology transfer to industry has reached an advanced stage of maturity. Recent demonstrations in industry-compatible devices now define the state of the art, both in terms of fabrication capabilities \cite{Bedecarrats2021,Zwerver2022,Bertrand2023,George2024,Neyens2024,jin2024,tomic2025,Koch2025,Ha2022,Ha2025} and qubit performance metrics \cite{Steinacker2024,nickl2025}.

This rapid progress, however, brings new challenges intrinsically linked to the increasing number of qubits. Many control and operation strategies that are effective for small arrays face severe limitations when extended to large-scale architectures. A central issue is the heating and power dissipation associated with high-frequency control signals, which constitute a bottleneck for high-fidelity operation in dense, large-scale systems \cite{Unseld2023,Unseld2025,champain2025,Capannelli2025}. In parallel, scaling up inevitably increases the number of control knobs per chip, including both DC and AC lines. To alleviate this overhead, shared-control architectures have been proposed, in which a reduced set of control lines is multiplexed across multiple qubits \cite{Borsoi2023,ivlev2025}.

These approaches are inherently sensitive to device variability. An increasing number of studies have highlighted the strong impact of disorder on spin qubits \cite{Martinez2022,Cifuentes2023,Martinez2024,Martinez2025,samadi2025,Valvo2025}, and how such variability can compromise the reliable operation of large-scale architectures \cite{Martinez2024,Martinez2025}. In systems with either intrinsic or artificial spin–orbit coupling, variability in the charge properties directly translates into variability of the spin properties, thereby amplifying the effects of disorder. This poses a critical challenge for spin manipulation in extended arrays, as conventional electric-dipole spin resonance (EDSR) techniques may struggle to deliver individualized high-frequency control to each qubit. When the number of available AC sources is limited, significant qubit-to-qubit variations in the Larmor frequency — both theoretically predicted \cite{Martinez2022,Martinez2025} and experimentally observed \cite{Hendrickx2023,john2024,seidler2025} — become increasingly difficult to compensate using Stark shifts alone.

A recent proposal for hole spin qubits offers an alternative strategy that cleverly circumvents the addressability issue and explicitly uses variability as a resource to manipulate the spin \cite{vanRiggelen-Doelman2024,wang2024}. This proposal leverages the strong intrinsic spin–orbit coupling of holes, which, in the presence of disorder, gives rise to spatially varying spin precession axes. Spin rotations are then carried out by electrically ``hopping'' the hole between neighboring QDs in a sparsely occupied array. Indeed, the spin precesses around different axes in each QD, so that properly timed transfers allow for arbitrary spin rotations \cite{wang2024}. This approach can be efficient down to small Larmor frequencies, allowing fast baseband control without the need for high-frequency driving. This strategy has thus the potential to mitigate both the power consumption constraints and the addressability challenges arising from qubit-to-qubit variability.

However, although spin manipulation by hopping is not compromised by disorder, it is fundamentally disorder-dependent. In this framework, disorder acts as a resource; yet, if variability is insufficient, the precession axes remain aligned and spin manipulation becomes ineffective. While attractive in the near term, this paradigm appears at odds with the long-term objective of continuously improving material quality, interface control, and device reproducibility \cite{Sammak2019,Scappucci2020,Scappucci2021,Meyer2023}.

In this work, we perform numerical simulations to assess the viability of spin manipulation by hopping as a function of disorder strength. We further propose an alternative strategy based on hopping between intentionally squeezed QDs. This approach retains the advantages of baseband control while removing the reliance on disorder, and remains robust against moderate variability. It therefore enables reliable spin manipulation even in the absence of disorder, offering improved prospects for scalable hole-spin quantum computing architectures.

The manuscript is organized as follows. Section \ref{sec:methodology} describes the simulation methodology. Section \ref{sec:circ} presents a statistical analysis of the reliability of spin manipulation by hopping in nominally circular QDs. Section \ref{sec:squeezed} then discusses the case of intentionally squeezed dots. Conclusions are summarized in Sec.~\ref{sec:conclusions}.

\section{Methodology} \label{sec:methodology}

\subsection{Definitions and protocol}

Spin manipulation by hopping leverages the difference between the precession axes in two neighboring quantum dots QD$_1$ and QD$_2$ to perform spin rotations by properly timed shuttles between the dots. Indeed, the spin precesses around a different axis when shuttled non-adiabatically from QD$_1$ to QD$_2$, which gives rise to a net spin rotation when brought back to QD$_1$ \cite{vanRiggelen-Doelman2024,wang2024}. This back-and-forth shuttle between the dots can be repeated as needed to prepare an arbitrary spin state in QD$_1$, and thus constitutes a building block for single-qubit operations.

More precisely, the precession axis of the hole spin is defined by the Larmor vector
\begin{equation} \label{eq:fL}
   h\boldsymbol{f}_L=\mu_B\hat{g}\boldsymbol{B}\,,
\end{equation}
where $h$ is the Plank constant, $\mu_B$ is the Bohr magneton, $\boldsymbol{B}=(B_x,B_y,B_z)$ is the magnetic field, and $\hat{g}$ is the $3\times 3$ gyromagnetic matrix of the hole spin \cite{Venitucci2018}. The Larmor frequency (spin precession speed) is $f_L=|\boldsymbol{f}_L|$. The Larmor vectors $\boldsymbol{f}_{L}^{(1)}$ of QD$_1$ and $\boldsymbol{f}_{L}^{(2)}$ of QD$_2$ make an angle $\alpha\in[0, 180\degree]$ such that:
\begin{equation}\label{eq:alpha}
    \cos\alpha=\frac{\boldsymbol{f}_{L}^{(1)}\cdot\boldsymbol{f}_{L}^{(2)}}{\left|\boldsymbol{f}_{L}^{(1)}\right|\left|\boldsymbol{f}_{L}^{(2)}\right|}\,.
\end{equation}
We assume that the Rashba and Dresselhaus spin-orbit interactions acting on the spin while it tunnels between the dots are negligible (as usual in strained Ge heterostructures \cite{Rodriguez2025}), and that shuttling is adiabatic with respect to charge but diabatic with respect to spin (namely, much faster than $1/f_L$ but much slower than $h/\tau$, with $\tau$ the charge tunnel coupling between the dots) \cite{Liu2026}. The spin is thus ``frozen'' while tunneling. The building block of the spin manipulation protocol consists of shuttling the hole from QD$_1$ to QD$_2$, then waiting for half a precession period $\Delta t_2=1/(2|\boldsymbol{f}_{L}^{(2)}|)$, shuttling the hole back to QD$_1$, and waiting for $\Delta t_1=1/(2|\boldsymbol{f}_{L}^{(1)}|)$ \cite{wang2024,Mauro2024strain}. After such a pulse sequence, the spin has undergone (in the laboratory frame) a $\pi$ rotation around $\boldsymbol{f}_{L}^{(2)}$ followed by a $\pi$ rotation around $\boldsymbol{f}_{L}^{(1)}$, which amounts to a rotation of angle $2\alpha$ around the axis $\boldsymbol{f}_{L}^{(2)}\times\boldsymbol{f}_{L}^{(1)}$ of the equatorial plane of the Bloch sphere of QD$_1$. The wait times and phase of the pulse sequence can be adjusted to reach any target point of that Bloch sphere. A $\pi$ rotation thus ideally requires $n_\pi=\lceil\pi/(2\alpha)\rceil$ back-and-forth hoppings, where $\lceil x\rceil$ is the smallest integer larger or equal to $x$. An angle $\alpha=90\degree$ actually allows single shot rotations to arbitrary states. The effective Rabi frequency of the spin rotations is thus
\begin{equation}\label{eq:fR}
    f_R=\frac{\alpha}{\pi}\overline{f}_L\,,
\end{equation}
where 
\begin{equation}
    2\overline{f}_L^{-1}=\left|\boldsymbol{f}_{L}^{(1)}\right|^{-1}+\left|\boldsymbol{f}_{L}^{(2)}\right|^{-1}
\end{equation}
is the average precession period.

\subsection{Device}
\begin{figure*}[t]
\centering
\includegraphics[width=0.98\textwidth]{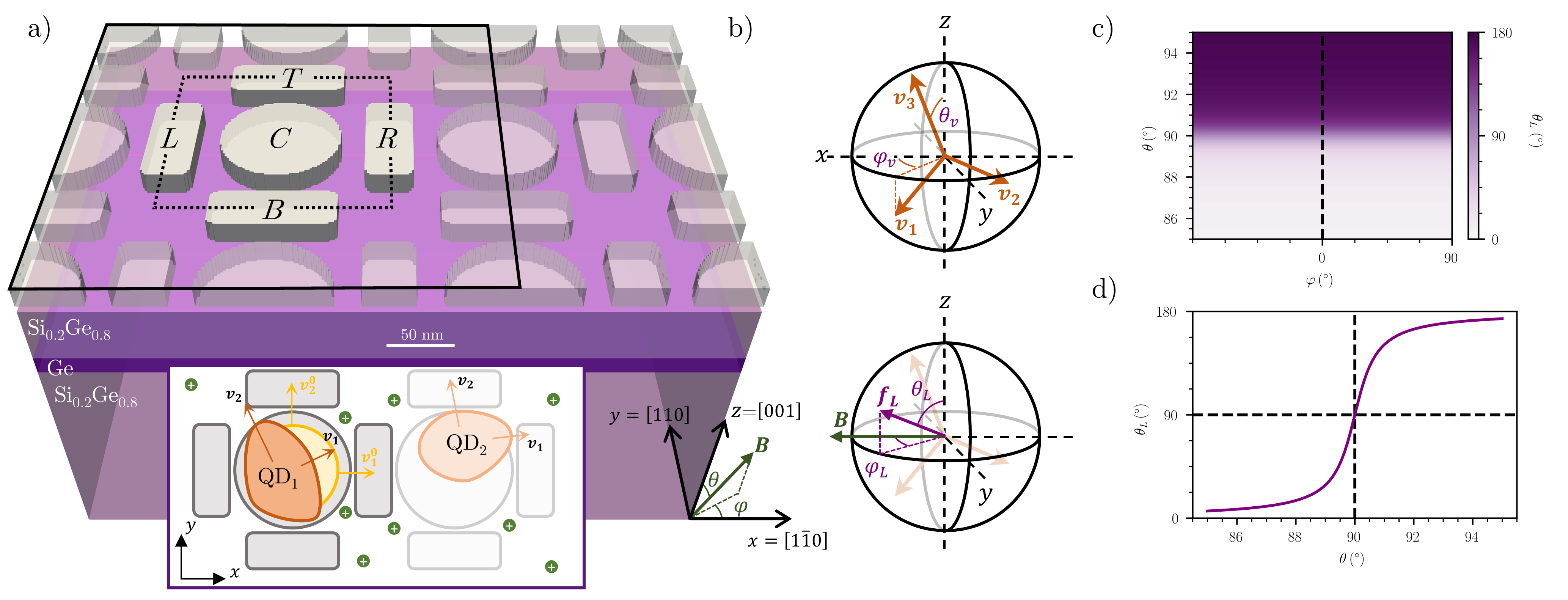}
\caption{a) The array of Ge hole spin qubits. SiGe is shown in light purple, Ge in dark purple, and the aluminium gates in gray. The dashed black line delimits the minimal unit cell of the array, and the solid black line the supercell considered for single QD calculations. The inset illustrates a top view of a double QD, where the two dots (orange) are deformed by the charged defects (green) with respect to the pristine QD (yellow). As a consequence, their magnetic axes $\boldsymbol{v}_1$, $\boldsymbol{v}_2$ (and $\boldsymbol{v}_3$) are scattered. b) Definition of the magnetic angles $(\theta_v,\varphi_v)$ describing the tilt of the magnetic axes $\{\boldsymbol{v}_1,\boldsymbol{v}_2,\boldsymbol{v}_3\}$ with respect to the device axes, and of the Larmor angles $(\theta_L,\varphi_L)$ characterizing the orientation of the Larmor vector. c) The Larmor angle $\theta_L$ as a function of the orientation of the magnetic field $\boldsymbol{B}=|\boldsymbol{B}|(\sin\theta\cos\varphi,\sin\theta\sin\varphi,\cos\theta)$ for a pristine QD ($V_C=-57.4$\,mV and all other gates grounded). d) The Larmor angle $\theta_L$ as a function of $\theta$ for $\varphi=0\degree$ (cut along the vertical dashed line in c)).}
\label{fig:dev}
\end{figure*}

Here we perform numerical simulations of spin manipulation by hopping in the periodic array of QDs shown in Fig. \ref{fig:dev}a. We consider a heterostructure with a 16 nm thick Ge well lying between a thick Ge$_{0.8}$Si$_{0.2}$ buffer and a 50\,nm thick Ge$_{0.8}$Si$_{0.2}$ barrier. We assume homogeneous residual strains in the buffer $\varepsilon_{xx}=\varepsilon_{yy}=+0.26\%$ \cite{Sammak2019}, which give rise to strains $\varepsilon_{xx}=\varepsilon_{yy}=\varepsilon_\parallel=-0.61\%$ and $\varepsilon_{zz}=\varepsilon_\perp=+0.45\%$ in the Ge well. The gate layout comprises a first level of barrier gates ($L$, $R$, $T$, $B$) that control the tunnel couplings between the dots, and a second level of plunger gates ($C$, diameter 100\,nm) that control the chemical potential of the dots. The two levels of aluminium gates are 20\,nm thick and are insulated from each other (and from the heterostructure) by 7\,nm of Al$_{2}$O$_{3}$. The gates are assumed to be connected to higher-lying metal (routing) levels by vias. The side of the unit cell of the array of QDs, outlined by the dashed black line in Fig. \ref{fig:dev}a, is 180\,nm. We account as in earlier works for the inhomogeneous strains imprinted by the thermal contraction of the gates at cryogenic temperatures \cite{Abadillo2022,Corley-Wiciak2023}. We focus in this work on the disorder due to charge traps in the gate stack, which we introduce as randomly distributed positive point charges at the GeSi/Al$_2$O$_3$ interface, with areal density $n_i$.

According to Eq.~\eqref{eq:alpha}, we can compute $\alpha$, $f_R$ and $n_\pi$ from the $g$ matrices of QD$_1$ and QD$_2$. We may, therefore, restrict our simulations to single dots. The minimal unit cell (dashed line) of Fig. \ref{fig:dev}a is, however, too small as periodic (or other) boundary conditions may give rise to strong spatial correlations in the disorder potential. We thus consider a $2\times 2$ supercell, outlined by the solid line in Fig. \ref{fig:dev}a, with a central dot of interest and three neighboring spectator dots (left empty). We further discuss in Appendix \ref{app:corr_double} the validity of this single dot strategy.

\subsection{Simulation workflow}

We follow the same workflow as in Refs. \cite{Martinez2022,Martinez2022-inhom,Abadillo2022,Martinez2024,Mauro2024,Martinez2025} for the numerical simulations. We mesh the whole structure with a product of non-homogeneous grids along $x$, $y$, and $z$. We compute strains with a finite-element discretization of the continuum elasticity equations (using the material parameters of Ref. \cite{Abadillo2022}). We solve Poisson's equation for the potential of the gates and charge traps with a finite-volume method, enforcing periodic boundary conditions along $x$ and $y$, and Neumann boundary conditions (zero normal electric field on the edges) along $z$. We next calculate the wave functions of the hole in these strains and potential with a finite-difference implementation of the four-band Luttinger-Kohn Hamiltonian on the same mesh \cite{Luttinger56,KP09}. As the hole is expected to be localized under the central $C$ gate, we break the periodic boundary conditions (which enables the application of an arbitrary magnetic field) and assume that the wave functions vanish at the edges of the simulation box. Finally, we make use of the $g$ matrix formalism \cite{Venitucci2018} to calculate the gyromagnetic matrix $\hat{g}$ of the QD from the ground doublet wave functions. 

We would like to emphasize that the $g$ matrix of a QD is not unique as it depends on the choice of magnetic axes (where to express $\boldsymbol{B}$), and on the choice of a (spin) basis set for the ground doublet subspace \cite{Venitucci2018}. The angle $\alpha$ is thus ill-defined if the spin bases used for QD$_1$ and QD$_2$ (both defined up to a unitary transform) are chosen randomly. We describe in Appendix \ref{app:spin_align} how to align these bases consistently with our assumption of diabatic spin shuttling between the dots.

We also remind that $\hat{g}$ is not necessarily symmetric:
\begin{equation}
    \hat{g}=
    \begin{pmatrix}
g_{11} & g_{12} & g_{13} \\
g_{21} & g_{22} & g_{23} \\
g_{31} & g_{32} & g_{33}
\end{pmatrix}\,.
\end{equation}
However, $\hat{g}$ can always be brought to a diagonal form with a singular value decomposition
\begin{equation}
    \hat{g}=\hat{U}\hat{g}_d\hat{V}^\intercal\,,
\end{equation}
where $\hat{g}_d=\textrm{diag}(g_1,g_2,g_3)$ and $\hat{U}$, $\hat{V}$ define unitary transformations of the spin and magnetic axes, respectively \footnote{Unlike the usual convention where the singular values are chosen positive, here we impose that the matrices $\hat{U}$, $\hat{V}$ have determinant $+1$ (direct orthonormal spin and magnetic axes). With this convention, the principal $g$ factors may be negative, as shown by the analytical expressions of Eq.~\eqref{eq:gHH}.}. The columns of $\hat{V}$ are the principal magnetic axes $\{\boldsymbol{v}_1,\boldsymbol{v}_2,\boldsymbol{v}_3\}$ of the QD. They are observable as the main axes of the ``peanut-shaped'' plots of the effective $g$ factor $g^*(\boldsymbol{b})=|\hat{g}\boldsymbol{b}|$ as a function of the orientation of the magnetic field $\boldsymbol{b}=\boldsymbol{B}/|\boldsymbol{B}|$. The columns $\{\boldsymbol{u}_1,\boldsymbol{u}_2,\boldsymbol{u}_3\}$ of $\hat{U}$ define a change of frame for $\boldsymbol{f}_L$ (a change of ``spin axes''), or equivalently a change of spin basis set for the doublet (in the original frame for $\boldsymbol{f}_L$) \cite{Venitucci2018}. With $\boldsymbol{B}=(B_1,B_2,B_3)$ in the principal magnetic axes $\{\boldsymbol{v}_1,\boldsymbol{v}_2,\boldsymbol{v}_3\}$, $h\boldsymbol{f}_L=\mu_B(g_1B_1,g_2B_2,g_3B_3)$ in the principal spin axes $\{\boldsymbol{u}_1,\boldsymbol{u}_2,\boldsymbol{u}_3\}$. Although $\hat{U}$ is not observable (as the spin basis is never explicit experimentally) \cite{Crippa2018}, the differences between the $\hat{U}$'s of neighboring QDs can contribute to $\alpha$ along with the differences between the $\hat{V}$'s and principal $g$ factors $(g_1,g_2,g_3)$.

For a pristine, quasi-circular QD ($V_T=V_B=V_R=V_L=0$\,V, $V_C<0$, no disorder), the magnetic axes can be aligned with the device axes: $\boldsymbol{v}_1^0=\boldsymbol{x}$, $\boldsymbol{v}_2^0=\boldsymbol{y}$, $\boldsymbol{v}_3^0=\boldsymbol{z}$ (we label hereafter the property $X$ of the pristine device as $X^0$) \footnote{Note that the in-plane effective $g$ factors are degenerate ($|g_1^0|=|g_2^0|$), so that the corresponding eigenvectors are actually ill-defined. We can however choose them aligned with $\boldsymbol{x}$ and $\boldsymbol{y}$.}. With a proper choice of spin basis set, the spin axes $\boldsymbol{u}_1^0=\boldsymbol{x}$, $\boldsymbol{u}_2^0=\boldsymbol{y}$, $\boldsymbol{u}_3^0=\boldsymbol{z}$ also align with the device axes. In this spin basis set, and in the $\{\boldsymbol{x},\boldsymbol{y},\boldsymbol{z}\}$ frame for $\boldsymbol{B}$, the $g$ matrix $\hat{g}^0$ of the pristine device is thus diagonal ($\hat{U}^0=\hat{V}^0=\hat{I}$). In the presence of disorder, the transformations $\hat{U}$ and $\hat{V}$ that diagonalize $\hat{g}$ are usually different from the identity. We characterize the rotation of the magnetic axes by the polar angle $\theta_v$ between $\boldsymbol{v}_3$ and $\boldsymbol{v}_3^0=\boldsymbol{z}$, and by the azimuthal $\varphi_v$ between the in-plane projection of $\boldsymbol{v}_1$ and $\boldsymbol{v}_1^0=\boldsymbol{x}$ (see Fig. \ref{fig:dev}b) \footnote{We define $\boldsymbol{v}_1$ as the magnetic axis ``closest to $\boldsymbol{v}_1^0=\boldsymbol{x}$'' ({\it i.e.}, with largest projection on $\boldsymbol{x}$).}. For a given magnetic field orientation, we also characterize the orientation of the Larmor vector by the polar angle $\theta_L$ between $\boldsymbol{f}_L$ and $\boldsymbol{z}$, and by the azimuthal angle $\varphi_L$ between the in-plane projection of $\boldsymbol{f}_L$ and $\boldsymbol{x}$. 

We assess the effects of disorder on the magnetic angles $(\theta_v,\varphi_v)$, the Larmor angles $(\theta_L,\varphi_L)$, and the rotation angle $\alpha$ between pairs of QDs. For that purpose, we collect statistics on at least 500 independent realizations of the single, disordered QD. To calculate $\alpha$, we use Eq.~\eqref{eq:alpha} on random pairs of dots  picked in this set of realizations. We discuss, moreover, the impact of variability on spin manipulation by analyzing the statistics of $f_R$ and $n_\pi$. We consider charge trap densities $n_i$ ranging from $10^{10}$\,cm$^{-2}$ to $10^{12}$\,cm$^{-2}$, which covers the experimental best and worst case scenarios. For each $n_i$, we adjust the $C$ gate voltage to achieve a target dot size in the reference pristine device with a uniform distribution of charges at the GeSi/Al$_2$O$_3$ interface (which accounts for the average effect of the traps on the confinement). For each property $f$ listed above, we choose the median $\overline{f}$ and the inter-quartile range (${\rm IQR}\equiv\mathcal{R}$) as statistical metrics for average behavior and spread, respectively (these metrics being more robust to outliers than the mean and standard deviation).

\section{Hopping between nominally circular dots} \label{sec:circ}

As explained above, spin manipulation by hopping relies on the difference of precession axes between QD pairs. Consequently, such a mechanism is not exploitable in an array of homogeneous, identical QDs. The vast majority of experimental demonstrations of hole spin qubits in germanium target nominally circular QDs \cite{Hendrickx2020,Hendrickx2021,Hendrickx2023,wang2024,seidler2025}. Still, all realizations show significant variability in the magnitude and orientational dependence of the in-plane $g$ factors, which is the fingerprint of broken symmetries that indeed yield to distinct precession axes. In the following, we discuss the requirements for spin manipulation by hopping, and provide statistical data on the role of disorder in achieving distinct precession axes for nominally circular QDs.

\subsection{Larmor vector in a pristine device}

As discussed in section \ref{sec:methodology}, the $g$ matrix of the pristine device can be made diagonal in the device axes set, with elements \cite{Michal2021,Martinez2022-inhom,Abadillo2022,Mauro2024strain}
\begin{subequations}
\label{eq:gHH}
\begin{align}
g_1^0&\approx +3q-\frac{6}{m_0\Delta_\mathrm{LH}}\left(\lambda\langle p_x^2\rangle-\lambda^\prime\langle p_y^2\rangle\right) \\
g_2^0&\approx -3q+\frac{6}{m_0\Delta_\mathrm{LH}}\left(\lambda\langle p_y^2\rangle-\lambda^\prime\langle p_x^2\rangle\right)\\
g_3^0&\approx6\kappa+\frac{27}{2}q-2\gamma_h\,,
\end{align}
\end{subequations}
where $\kappa=3.41$ and $q=0.06$ are the isotropic and cubic Zeeman parameters of Ge, $m_0$ is the free electron mass, $\Delta_\mathrm{LH}$ is the heavy-hole/light-hole bandgap of the QD, $\lambda=\kappa\gamma_3-2\eta_h\gamma_2\gamma_3\approx -0.38$, and $\lambda^\prime=\kappa\gamma_3-2\eta_h\gamma_3^2\approx -7.15$, with $\gamma_2=4.24$ and $\gamma_3=5.69$ the Luttinger parameters of bulk Ge. The squared momentum operators $\langle p_x^2\rangle$ and $\langle p_y^2\rangle$ are averaged over the ground-state heavy-hole envelope. The factors $\gamma_h\approx 2.62$ and $\eta_h\approx 0.41$ depend on vertical confinement \cite{Michal2021}. The above equations account for the effects of structural, electrical and magnetic confinement on the gyromagnetic response of the hole. For a large, quasi-circular quantum dot ($\langle p_x^2\rangle=\langle p_y^2\rangle$), the principal in-plane $g$ factors along $\boldsymbol{x}$ and $\boldsymbol{y}$ are $g_1^0=-g_2^0\equiv g_\parallel^0\approx 3q=0.18$, while the out-of-plane $g$ factor along $z$ is $g_3^0\equiv g_\perp^0\approx 15$. The gyromagnetic response is thus highly anisotropic, $g_\perp^0/g_\parallel^0$ being of the order of one hundred. As a consequence, the Larmor vector $\boldsymbol{f}_L^0=(g_\parallel^0 B_x, -g_\parallel^0 B_y, g_\perp^0 B_z)$ locks onto the $z$ axis once the magnetic field slightly goes out-of-plane \cite{Mauro2024strain,Mauro2025}. This is illustrated in Fig. \ref{fig:dev}c, which plots the Larmor angle $\theta_L^0$ as a function of the orientation of the magnetic field  $\boldsymbol{b}=\boldsymbol{B}/|\boldsymbol{B}|=(\sin\theta\cos\varphi,\sin\theta\sin\varphi,\cos\theta)$. For a magnetic field only one degree out-of-plane $\boldsymbol{f}_L^0$ is already almost parallel to $\boldsymbol{z}$.

\begin{figure*}[t]
\centering
\includegraphics[width=0.98\textwidth]{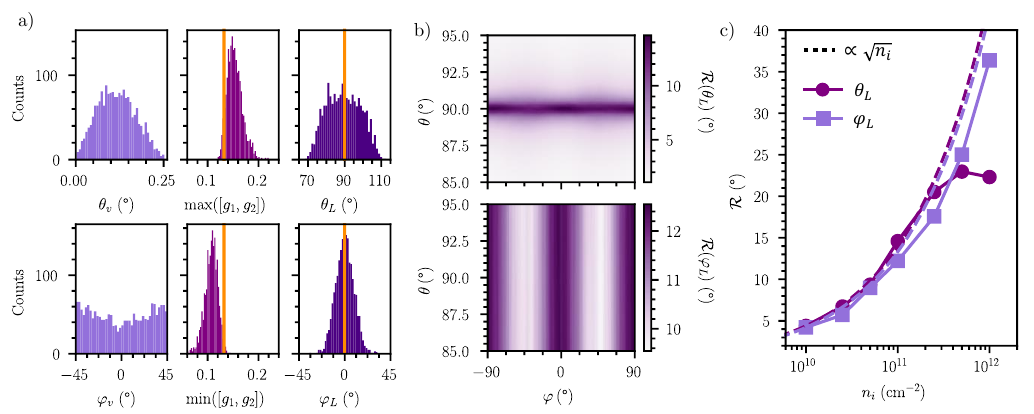}
\caption{Scattering of Larmor angles in the presence of disorder in nominally circular QDs with size $\ell_c^0=20$\,nm. a) Histogram of the magnetic angles $(\theta_v,\varphi_v)$, principal $g$ factors $(g_1,g_2)$, and Larmor angles $(\theta_L,\varphi_L)$ in disordered QDs with charge trap density $n_i=10^{11}$\,cm$^{-2}$. The orange lines are the values for the pristine device. b) Inter-quartile range ($\mathcal{R}$) of the Larmor angles as a function of the magnetic field orientation. c) Dependence of the inter-quartile range of the Larmor angles on $n_i$. The dashed lines are guides to the eye for the expected $\propto\sqrt{n_i}$ dependence \cite{Martinez2022}. In panels a) and c), $\theta_L$ and $\varphi_L$ are calculated for a magnetic field $\boldsymbol{B}\parallel\boldsymbol{x}$.}
\label{fig:angles}
\end{figure*}

Spin manipulation by hopping can not be achieved in the pristine device, as two identical circular QDs have same $\boldsymbol{f}_L^{(1)}=\boldsymbol{f}_L^{(2)}=\boldsymbol{f}_L^{0}$, so that $\alpha=0$. It is, therefore, a mechanism enabled by disorder, as we discuss hereafter. Here we consider charge traps at the SiGe/Al$_2$O$_3$ interface as the main source of disorder. In real devices, stray fields due to fanout (contact) lines, uniaxial \cite{Mauro2024strain,seidler2025} or long-range strains due to cross-hatch patterns \cite{Nigro2025, pena2026} may also break the circular symmetry of the QD. We neglect these effects that are strongly depend on device layout and growth conditions. We address in Appendix \ref{app:doubleDot} the impact of detuning in a double QD device, which also breaks the circular symmetry, but only makes a minor contribution to $\alpha$.

\subsection{Spin manipulation in disordered QDs} \label{sec:circ_dis}

Achieving $\alpha\neq 0$ requires QDs with different $\hat{g}$ so that $\boldsymbol{f}_L^{(1)}\nparallel\boldsymbol{f}_L^{(2)}$. These differences may arise from the mismatch of the principal $g$ factors (different $\hat{g}_d$), and from the mismatch of the spin ($\hat{U}$) and magnetic ($\hat{V}$) axes. As hinted in Refs. \cite{Martinez2025,Valvo2025}, disorder can indeed result in such an imbalance between neighboring QDs as spin-orbit interactions couple the disorder-dependent position, size, and shape of the quantum dots to their spins \cite{Michal2021,Martinez2022-inhom,Abadillo2022}.

We first plot in Fig. \ref{fig:angles}a the distribution of the magnetic angles $(\theta_v,\varphi_v)$ for a dataset of 2\,000 QDs with nominal size (pristine device) $\ell_c^0=\ell_x^0=\ell_y^0=20$\,nm ($\ell_\alpha=\sqrt{\braket{\alpha^2}-\braket{\alpha}^2}$), and a charge trap density $n_i=10^{11}$\,cm$^{-2}$. The principal $g$ factors of the pristine QD are $g_\parallel^0=0.13$ and $g_\perp^0=13.74$. In the presence of disorder, the vertical magnetic axis $\boldsymbol{v}_3$ remains tightly locked onto $\boldsymbol{z}$ by the large gyromagnetic anisotropy ($\theta_v\lesssim 0.25\degree$), and the principal $g$ factor $g_3\approx g_\perp^0$ (not shown) is robust. As discussed below, this small $\theta_v$ leaves, nevertheless, visible fingerprints on the spin dynamics. It results from the coupling between the in- and out-of-plane motions of the hole in the disordered QDs, and from the inhomogeneous shear strains imprinted by differential thermal contraction of the materials \cite{Martinez2022-inhom,Abadillo2022}. The degeneracy between the effective in-plane $g$ factors of the pristine QD ($|g_1^0|=|g_2^0|$) is lifted by the charge traps ($|g_1|\ne|g_2|$) due to the variation of the in-plane QD size and shape [see Eq. \eqref{eq:gHH}]. As a consequence of this original degeneracy, the in-plane magnetic axes $\boldsymbol{v}_1$ and $\boldsymbol{v}_2$ get (almost) uniformly scattered by disorder \footnote{As discussed in Ref. \cite{Martinez2025}, the slight preferential alignment along $\varphi\pm 45\degree$ results from the weaker screening of the charges located along these directions by the gates.}.

\begin{figure}[t]
\centering
\includegraphics[width=0.98\columnwidth]{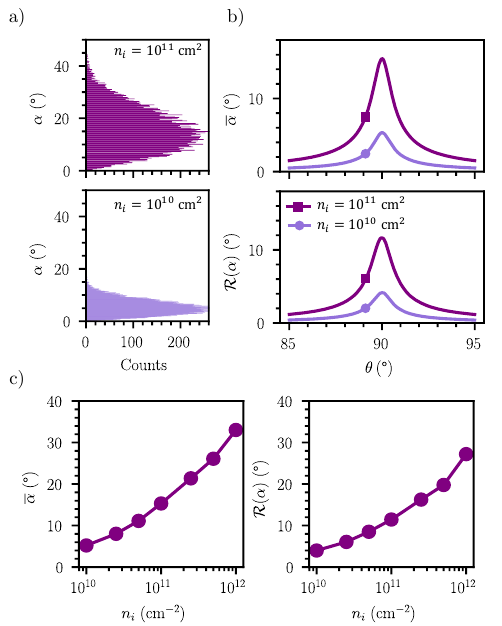}
\caption{Statistics of the angle between the Larmor vectors of pairs of nominally circular QDs with size $\ell_c^0=20$\,nm. a) Histogram of the angle $\alpha$ between the Larmor vectors of QD pairs ($\boldsymbol{B}\parallel\boldsymbol{x}$), for charge trap densities $n_i=10^{11}$\,cm$^{-2}$ (top panel) and $n_i=10^{10}$\,cm$^{-2}$ (bottom panel). b) Dependence of the median $\overline{\alpha}$ (top panel) and inter-quartile range $\mathcal{R}(\alpha)$ (bottom panel) on the magnetic field angle $\theta$ for $\varphi=0\degree$. The trends are rather independent of $\varphi$. c) Dependence of $\overline{\alpha}$ and $\mathcal{R}(\alpha)$ on $n_i$ ($\boldsymbol{B}\parallel\boldsymbol{x}$).}
\label{fig:angles_pairs}
\end{figure}

The fluctuations of the principal $g$ factors $g_1$, $g_2$, of the magnetic angles $\theta_v$, $\varphi_v$, and the rotations of the spin axes (discussed in Appendix \ref{app:corr_spin}) give rise to a variability of the precession axis of the spin. In the presence of disorder, all matrix elements of $\hat{g}$ may actually be non-zero. Let us assume $\boldsymbol{B}\parallel\boldsymbol{x}$ as an example; the Larmor vector then reads
\begin{equation}
    \boldsymbol{f}_L=\frac{\mu_B|\boldsymbol{B}|}{h}(g_{11},g_{21},g_{31})\,.
\end{equation}
In the pristine device, $g_{21}=g_{31}=0$, so that $\boldsymbol{f}_L^{0}\parallel\boldsymbol{x}$. Therefore, a rotation of the magnetic and/or spin axes, responsible for a finite $g_{21}$ and $g_{31}$, is needed to tilt the precession axis. The Larmor angle $\theta_L$ (see Fig. \ref{fig:dev}b for the definition) is actually controlled by $g_{31}$, and the Larmor angle $\varphi_L$ by $g_{21}$. As hinted above, $g_{31}\propto\braket{p_x p_z},\,\braket{\varepsilon_{xz}}$ essentially results from the coupling between the in- and out-of-plane motions of the hole in the non-separable potential of the device \cite{Martinez2022-inhom}, and from the shear strains $\varepsilon_{xz}$ imprinted by differential thermal contraction \cite{Abadillo2022} (but is actually dominated by the latter). On the other hand, $g_{21}\propto\braket{p_x p_y},\,\braket{\varepsilon_{xy}}$ primarily results from the deformations of the dot (ellipticity/rotation), and, to a lesser extent, from the in-plane shear strains $\varepsilon_{xy}$ \cite{Michal2021,Abadillo2022,Mauro2024strain}. 

We plot in Fig. \ref{fig:angles}a the distributions of $\theta_L$ and $\varphi_L$ for $n_i=10^{11}$\,cm$^{-2}$, and in Fig. \ref{fig:angles}b the IQR of both angles as a function of the magnetic field orientation. The median (not shown) follows the same behavior as in the pristine device (Fig. \ref{fig:dev}c,d). The normalized precession vector $\boldsymbol{u}_L=\boldsymbol{f}_L/|\boldsymbol{f}_L|=(\sin\theta_L\cos\varphi_L,\sin\theta_L\sin\varphi_L,\cos\theta_L)$ closely follows a von-Mises-Fisher distribution, the analog of a Gaussian distribution on the unit sphere. Interestingly, even though $\theta_v$ is extremely small, the fluctuations of $\theta_L$ can be as large as those of $\varphi_L$ for in-plane magnetic fields. Indeed, the off-diagonal element $g_{31}$ is too small with respect to $g_{33}\approx g_\perp^0$ to tilt $\boldsymbol{v}_3$, yet it can be comparable in magnitude to $g_{11}$, and thus has a significant impact on $\boldsymbol{f}_L$ when $B_z\approx 0$. Yet, $\mathcal{R}(\theta_L)$ rapidly vanishes when $\theta\lessgtr 90\degree$ because, as discussed previously, the Larmor vector rapidly locks onto $\boldsymbol{z}$. Achieving heterogeneous precession axes requires, therefore, a careful in-plane alignment of $\boldsymbol{B}$ \cite{wang2024}.

\begin{figure}[t]
\centering
\includegraphics[width=0.98\columnwidth]{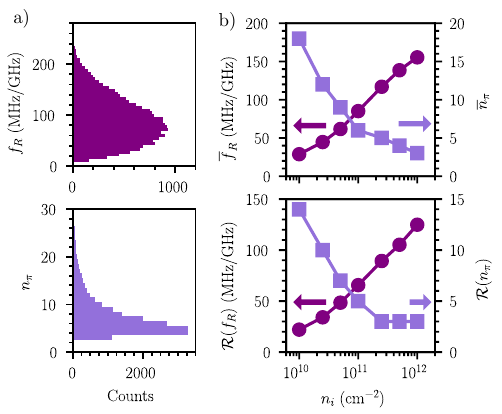}
\caption{Statistics on the Rabi frequencies and number of shuttling pulses. a) Histogram of the Rabi frequency $f_R$ and number of shuttling pulses $n_\pi$ required to perform a $\pi$ rotation of the spin for disordered QDs with charge trap density $n_i=10^{11}$\,cm$^{-2}$. b) Median (top panel) and inter-quartile range (bottom panel) of $f_R$ and $n_\pi$ as a function of  $n_i$. In all panels, $\ell_c^0=20$\,nm and $\boldsymbol{B}\parallel\boldsymbol{x}$.}
\label{fig:fR}
\end{figure}

The magnitude of the IQRs of both $\theta_L$ and $\varphi_L$ is roughly $12\degree$ for $\boldsymbol{B}\parallel\boldsymbol{x}$ and $n_i=10^{11}$\,cm$^{-2}$, and is weakly dependent on the orientation of an in-plane magnetic field. Importantly, $\mathcal{R}$ increases as $\propto\sqrt{n_i}$ for small $n_i$ \cite{Martinez2022}, as shown in Fig. \ref{fig:angles_pairs}c. It is barely $5\degree$ for $n_i=10^{10}$\,cm$^{-2}$, but reaches $\mathcal{R}(\theta_L)\approx24\degree$ and $\mathcal{R}(\varphi_L)\approx37\degree$ for $n_i=10^{12}$\,cm$^{-2}$. These results highlight the tight relation between the strength of disorder and the heterogeneity of the precession axes in the QDs. We further analyze this relation by collecting statistics on the angle $\alpha$ between the precession axes of pairs of QDs.  

\begin{figure}[t]
\centering
\includegraphics[width=0.98\columnwidth]{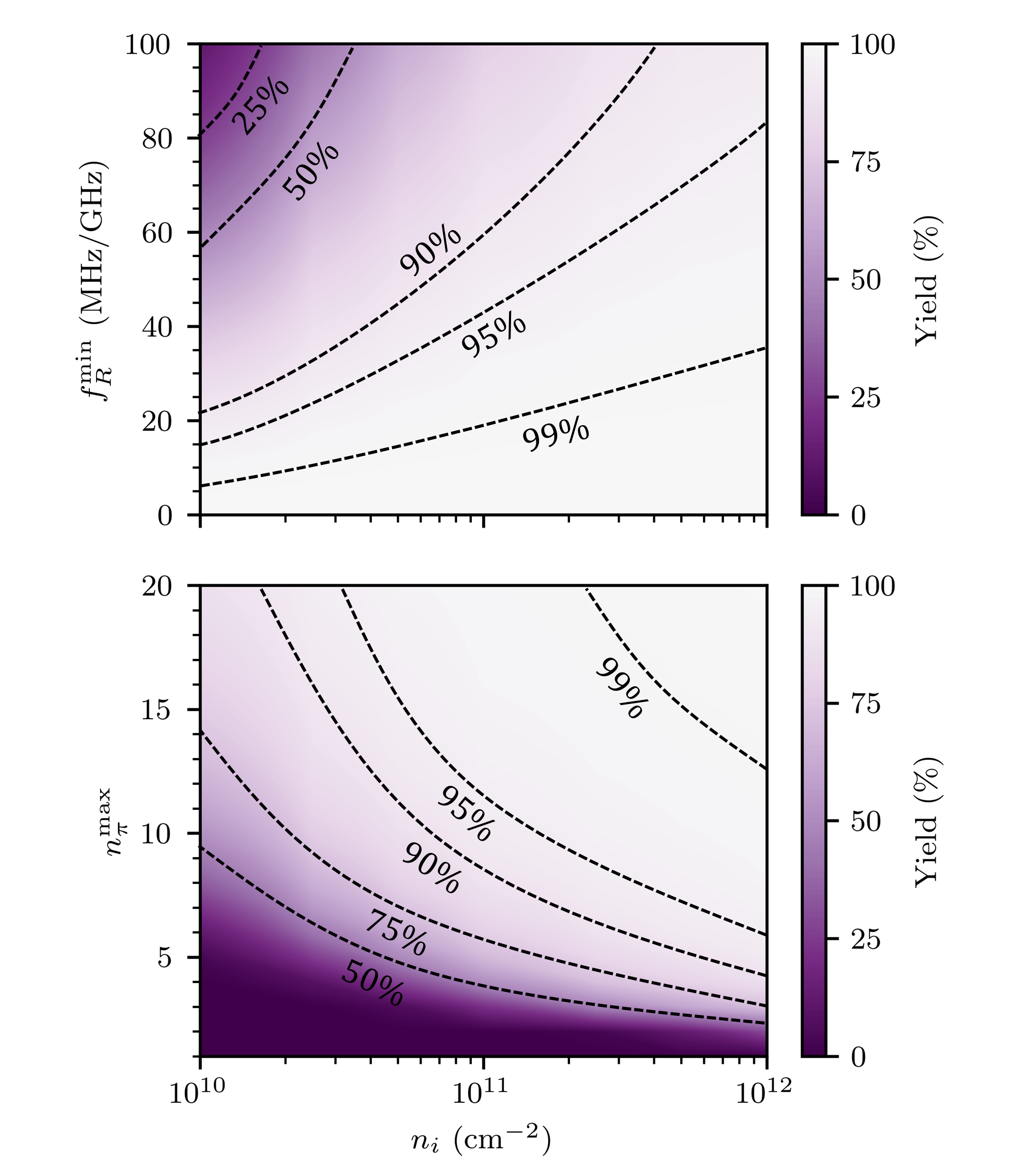}
\caption{Percentage of functional quantum dot pairs (yield) as a function of the charge trap density $n_i$. The functionality condition is $f_R>f_R^{\rm min}$ for the top panel, and $n_\pi<n_\pi^{\rm max}$ for the bottom panel. For all $n_i$, $\ell_c^0=20$\,nm and $\boldsymbol{B}\parallel\boldsymbol{x}$.}
\label{fig:fR_small}
\end{figure}

Fig. \ref{fig:angles_pairs}a shows the histogram of $\alpha$ for a dataset of 20\,000 independent QD pairs, $\boldsymbol{B}\parallel\boldsymbol{x}$ and $n_i=10^{11}$\,cm$^{-2}$. The shape of this histogram, which peaks at finite $\alpha$, is characteristic of the angle between two unit vectors $\boldsymbol{u}_L^{(1)}$ and $\boldsymbol{u}_L^{(2)}$ drawn from a von-Mises-Fisher distribution \footnote{The histogram count is $\approx 0$ when $\alpha\to 0$ as the surface $2\pi|\sin\alpha|d\alpha$ between shells $\alpha$ and $\alpha+d\alpha$ on the unit sphere also tends to zero (the density of configurations with $\alpha\to 0$ vanishes).}. The median $\overline{\alpha}$ and IQR $\mathcal{R}(\alpha)$ are plotted as a function of the density of charge traps in Fig. \ref{fig:angles_pairs}c. As expected, $\overline{\alpha}$ tends to zero when $n_i\to 0$ (no disorder, identical QDs) and increases with increasing $n_i$. As a consequence, $\mathcal{R}(\alpha)$ is comparable to $\overline{\alpha}$. The median angle between QD pairs can be as large as $\approx35\degree$ for $n_i=10^{12}$\,cm$^{-2}$, but barely reaches $6\degree$ for $n_i=10^{10}$\,cm$^{-2}$. This illustrates the main strength and weakness of spin manipulation by hopping in nominally circular QDs: it leverages disorder as an asset rather than a threat, yet it requires large levels of disorder to be effective on average. Moreover, the distribution always features a tail of QD pairs with small $\alpha$ (even for large disorders), whose manipulation is slow. We explore these limitations hereafter.

In Fig. \ref{fig:angles_pairs}b we plot both $\overline{\alpha}$ and $\mathcal{R}(\alpha)$ as a function of the magnetic field orientation. Again, increasing $B_z$ rapidly aligns all precession axes along $\boldsymbol{z}$ ($\mathcal{R}(\theta_L)$ drastically decreases in Fig. \ref{fig:angles}b), and consequently $\alpha$ vanishes. Achieving significant $\alpha$ therefore requires tight in-plane alignment of $\boldsymbol{B}$. The statistics are almost independent of the azimuthal angle $\varphi$ of the magnetic field owing to the circular symmetry of the pristine QD. 

As explained in the Methodology section, both the effective Rabi frequency $f_R$ and the number of shuttling pulses $n_\pi$ required to perform a $\pi$ rotation are proportional to $\alpha$. In Fig. \ref{fig:fR}a we plot the distribution of $f_R$ and $n_\pi$ for $\boldsymbol{B}\parallel\boldsymbol{x}$ and $n_i=10^{11}$\,cm$^{-2}$. $f_R$ is computed at the magnetic field $|\boldsymbol{B}|=0.54$\,T such that the Larmor frequency of the pristine device is $f_L^0=1$\,GHz. It is expected to be proportional to $f_L^0$, and is thus given in MHz/GHz. $f_R$ inherits the distribution of $\alpha$, with a tail of fast pairs of QDs with large angles that consequently require fewer shuttling pulses $n_\pi$. There is nonetheless a long tail of devices in the distribution of $n_\pi$ that require $>10$ back-and-forth hoppings, which correspond to those pairs of QDs with small $\alpha$. We emphasize, though, that the median ratio $\overline{f_R}/f_L^0\approx85$ MHz/GHz is actually much larger than for traditional, intra-dot EDSR (owing to the much larger, effective inter-dot dipole) \cite{Martinez2022-inhom,Abadillo2022}. This makes spin manipulation by hopping efficient down to small Larmor frequencies $<200$\,MHz, indeed enabling baseband control. In Fig. \ref{fig:fR}b we plot the median and IQR of $f_R$ and $n_\pi$ as a function of $n_i$. These data clearly highlight once again the limitations of this spin manipulation technique: the rotations become slower and require more shuttles with decreasing $n_i$. For high quality interfaces, namely $n_i\lesssim 5\times 10^{10}$\,cm$^{-2}$, the median $n_\pi$ is expected to be larger than 10, and the median Rabi frequency to be below 75\,MHz/GHz. The experiment of Ref. \cite{wang2024} was actually run with a magnetic field of 25\,mT and $f_L$'s in the $30-90$\,MHz range, yielding net $f_R$'s $\approx 10-20$ times smaller.

In Fig. \ref{fig:fR_small} we plot the percentage of devices with a Rabi frequency larger than a target minimum value ($f_R^{\rm min}$), and the percentage of devices with $n_\pi$ smaller than a target maximum number of pulses ($n_\pi^{\rm max}$), as a function of the level of disorder $n_i$. The results further emphasize the complications arising from the skewed distributions discussed above. A non-negligible fraction of qubits display slow $f_R$ at low disorder. At $n_i=10^{12}$\,cm$^{-2}$, only 0.4\% of the devices feature $f_R<20$\,MHz/GHz, yet for $n_i=10^{10}$\,cm$^{-2}$, this fraction almost reaches 9\%. For a typical Larmor frequency of 50 MHz, $f_R=20$\,MHz/GHz translates into a $\pi$ pulse gate time of 1 $\mu$s, which is already a significant fraction of typical dephasing times ($T_2^*\approx 5$\,$\mu$s at $|\boldsymbol{B}|=25$\,mT in Ref. \cite{wang2024}) \footnote{We want to highlight that both $f_R$ and $T_2^*$ are proportional to $B$, thus the relevance of normalizing $f_R$ per GHz of Larmor frequency.}. Note that the slowest qubits, together with those with a shorter coherence, are the ones limiting the quality factor of a quantum processor \cite{Martinez2022,Martinez2025}. Similarly, the number of qubits that can be manipulated with at most $n_\pi^{\rm max}$ shuttling pulses drastically decreases with reducing $n_i$. As an example, 98\% of the devices can be manipulated with $n_\pi\leq 10$ for $n_i=10^{12}$\,cm$^{-2}$, yet only 53\% for $n_i=10^{10}$\,cm$^{-2}$. Moreover, systematically operating with a very limited number of pulses $n_\pi^{\rm max}<5$ is out of reach for 31\% of the devices for $n_i<10^{11}$\,cm$^{-2}$, and it is not possible for 8\% of the qubits even at $n_i=10^{12}$\,cm$^{-2}$.

The results discussed in this section illustrate that the implementation of spin manipulation by hopping in clean, little defective devices may be challenging. While being a viable option for few-qubit experiments, with the robust and efficient operation of larger arrays being strongly intertwined with the improvement of material and interface qualities, it is hard to foresee a scalable roadmap with spin manipulation relying on large disorder. Alternatively, engineering the $\hat{g}$ matrix of the QD pairs to achieve $\alpha^0\neq 0\degree$ is an appealing option to make spin manipulation by hopping both independent on and resilient to disorder. We propose to do so by squeezing the QDs in the next section.

\section{Hopping between squeezed dots} \label{sec:squeezed}

\begin{figure*}[t]
\centering
\includegraphics[width=0.98\textwidth]{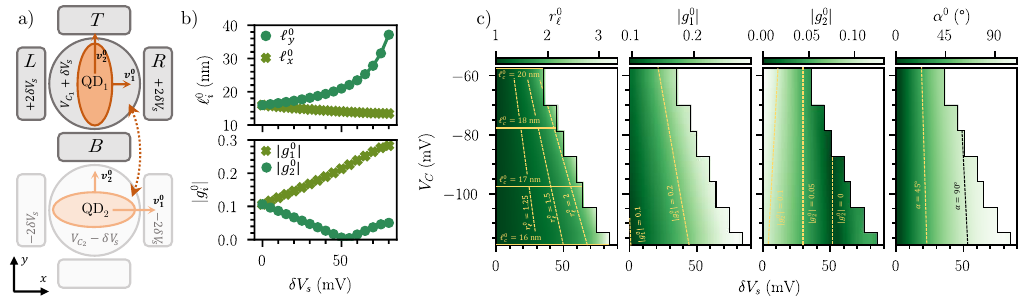}
\caption{Hopping between squeezed dots in a pristine device. a) Schematic representation of hopping between orthogonally squeezed QDs. The dots (in orange) are squeezed electrostatically with opposite voltages on the barrier gates (in gray). A pair of orthogonally squeezed QDs shares the same magnetic axes, yet the effective $g$ factors are swapped along $\boldsymbol{x}$ and $\boldsymbol{y}$. b) Dependence of the in-plane sizes $\ell_x^0$ and $\ell_y^0$ of single QDs, and of the in-plane principal $g$ factors $g_1^0$ and $g_2^0$ on the squeezing voltage $V_{L,R}=-V_{T,B}=\delta V_s$. The data are calculated for $V_C=-117.44$\,mV ($\ell_x^0=\ell_y^0=\ell_c^0=16$\,nm when $\delta V_s\to 0$). c) Squeezing ratio $r_\ell^0$, principal $g$ factors $|g_1^0|$ and $|g_2^0|$ of QD$_1$, and angle $\alpha^0$ between the precession axes of QD$_1$ and QD$_2$ ($\boldsymbol{B}\parallel\boldsymbol{x}+\boldsymbol{y}$) as a function of the gate voltages. The blank areas highlight the gate voltages range where squeezing is too strong and the dot splits into a strongly-coupled double QD.}
\label{fig:sq_ideal}
\end{figure*}

The dots can be squeezed to take an elliptical (rather than circular) shape \cite{Bosco2021}, the squeezing axis being the minor axis (strong confinement direction) of the ellipse. The effects of the reduced symmetry of squeezed dots on the gyromagnetic response are twofold: this locks the in-plane magnetic axes $\boldsymbol{v}_1$, $\boldsymbol{v}_2$ along the major and minor axes of the ellipse, and this breaks the degeneracy \cite{Mauro2024,Valvo2025} between the principal in-plane $g$ factors $g_1$, $g_2$. By squeezing neighboring dots along orthogonal axes (see Fig. \ref{fig:sq_ideal}a), we can engineer pairs of QDs with same magnetic axes but different principal $g$ factors. This results in distinct precession axis for the two dots (if $\boldsymbol{B}$ is not parallel to a magnetic axis), which can be exploited to manipulate the spin by hopping. This strategy is deterministic in the sense that it achieves finite $\alpha$'s even in pristine devices, and thus does not rely on disorder. If the squeezing and $g$ factor engineering are robust against disorder, such a strategy indeed represents a disorder-resilient yet disorder-independent spin manipulation technique. 

This proposal shares part of the underlying physics with the gapless spin qubit of Ref. \cite{Russ2025}. The gapless spin qubit manipulation indeed relies on diabatic squeezing along perpendicular in-plane directions to break the $g$ factor degeneracy and induce spin rotations along orthogonal axes. The idling state is nonetheless a circular QD that is electrically tuned to $g_1=g_2\approx0$ to suppress all dynamics, and the diabatic pulses are applied within the same QD (there is no hopping involved). However, such an operating point remains to be experimentally demonstrated; and our attempts to reach $g_1=g_2=0$ in the device of Fig. \ref{fig:dev}a have been unsuccessful, as the required lateral electric fields are too strong and the QD escapes the Ge well before achieving the target condition \footnote{We have considered a device with top GeSi barrier thickness $H=50$ nm. Tighter electrical confinements can be achieved with $H<50$ nm, which may ease the achievement of the idling condition $g_1=g_2=0$.}. Therefore, our proposal gives up the degenerate idling state, yet it exploits the $g$ factor modulation and axis locking by squeezing proposed in Refs. \cite{Mauro2024, Russ2025,Valvo2025} to engineer a disorder-independent version of the hopping spins of Ref. \cite{wang2024}. Baseband hopping between squeezed QDs shall minimize power consumption and is compatible with scalable architectures with minimal overhead (see proposal in Appendix \ref{app:sq_architecture}). We highlight, however, that the inter-dot shuttling pulses could alternatively be substituted by intra-dot squeezing pulses, at the likely cost of a higher power consumption (see Appendix \ref{app:intradot}).

In the following sections, we describe in more detail the physics of spin manipulation by hopping between pairs of orthogonally squeezed QDs. We squeeze the QDs electrostatically by biasing the $L$, $R$, $T$, $B$ gates asymmetrically (see Fig. \ref{fig:dev}a). We first illustrate the mechanism for pristine devices in Sec. \ref{sec:sq_id}, and then study the resilience of this mechanism to disorder in Sec. \ref{sec:sq_disorder}. We also discuss more scalable, geometrical approaches to squeezing, where the shape of the QDs is engineered by the gate layout rather than by biasing multiple gates, in Appendix \ref{app:sq_architecture}.

\subsection{Spin manipulation in pristine devices} \label{sec:sq_id}

We consider pristine dots squeezed along either $x$ ($\ell_y^0>\ell_x^0$) or $y$ ($\ell_x^0>\ell_y^0$). As mentioned above, the reduced symmetry locks the in-plane magnetic axes along $\boldsymbol{x}$ and $\boldsymbol{y}$, and breaks the degeneracy of the in-plane $g$ factors (as evidenced by Eqs.~\eqref{eq:gHH} when $\braket{p_x^2}\propto \ell_x^{-2}$ is different from $\braket{p_y^2}\propto \ell_y^{-2}$). These effects can be exploited to perform spin manipulation by hopping between two orthogonally squeezed QDs (see Fig. \ref{fig:sq_ideal}a). We assume here that QD$_1$ is squeezed along $x$, and that QD$_2$ is identically squeezed along $y$ (but the opposite would be equivalent). If $g_1^0$ and $g_2^0$ are the principal $g$ factors of QD$_1$ along $\boldsymbol{x}$ and $\boldsymbol{y}$, then Eqs.~\eqref{eq:gHH} show that the principal $g$ factors of QD$_2$ are $-g_2^0$ and $-g_1^0$ along the same axes.

Let us now apply a purely in-plane magnetic field \footnote{The out-of-plane $g$ factor barely changes with squeezing, so the same tight in-plane alignment of $\boldsymbol{B}$ discussed for the circular QDs applies to squeezed QDs as well.}. The Larmor vectors of the QD pair then read
\begin{subequations}
\label{eq:pristinelarmors}
\begin{align}
\boldsymbol{f}_L^{(1)} &= \frac{\mu_B }{h}(g_1^0B_x,g_2^0B_y,0) \\
\boldsymbol{f}_L ^{(2)} &= -\frac{\mu_B}{h}(g_2^0B_x,g_1^0B_y,0)\,,
\end{align}
\end{subequations}
so that the angle between the spin precession axes is
\begin{equation}\label{eq:alpha_sq}
\cos\alpha^0=\frac{-g_1^0g_2^0(B_x^2+B_y^2)}
{\sqrt{(g_1^0B_x)^2+(g_2^0B_y)^2} \, 
 \sqrt{(g_2^0B_x)^2+(g_1^0B_y)^2}}\,.
\end{equation}
This expression highlights that $\alpha^0$ is finite even in the pristine devices as long as $\boldsymbol{B}$ is not parallel to the magnetic axes $\boldsymbol{x}$ or $\boldsymbol{y}$. The optimal (largest) $\alpha$ is actually achieved for $B_x=B_y$ ($\theta=90\degree,\varphi=45\degree$), where $\cos\alpha^0=-2g_1^0g_2^0/((g_1^0)^2+(g_2^0)^2)$. With the magnetic axes locked, manipulation by hopping between squeezed QDs indeed benefits from the broken in-plane degeneracy ($g_1^0\neq g_2^0$).

In our single dot simulations, we squeeze the QDs by applying opposite voltages to the $L$, $R$ and $T$, $B$ gates ($V_{L,R}=-V_{T,B}=\delta V_s$) at constant $C$ gate bias. Orthogonal squeezing in a pair of neighboring dots (see Fig.~\ref{fig:sq_ideal}a) can be achieved by applying a $+2\delta V_s$ shift to the $L$, $R$ gates of QD$_1$, and a $-2\delta V_s$ shift to the same gates of QD$_2$, along with $\pm\delta V_s$ shifts on the $C$ gates. This is roughly equivalent to the above single dot squeezing scheme (the differences of potential between the plunger and barrier gates of each dot being the same). Conveniently, this leaves the central B gate available to control the tunnel coupling between the dots (see Appendix \ref{app:tc}). In Fig. \ref{fig:sq_ideal}b we illustrate how the sizes $\ell_x^0$ and $\ell_y^0$, as well as the effective $g$ factors $|g_1^0|$ and $|g_2^0|$ of the single dot vary with the squeezing voltage $\delta V_s$. The size of the QD is $\ell_c^0=\ell_x^0=\ell_y^0=16$\,nm in the absence of squeezing ($V_C=-117.44$\,mV and $\delta V_s\to 0$). These results demonstrate that the dot is effectively squeezed ($\ell_x^0<\ell_c^0$ and $\ell_y^0>\ell_c^0$ if $\delta V_s>0$), and, by doing so, that the principal $g$ factors are modulated as expected from Eqs.~\eqref{eq:gHH}. Interestingly, $g_2^0$ changes sign at $\delta V_s\approx 50$\,mV \cite{Mauro2024}. There is actually experimental evidence of this behavior in Ref. \cite{wang2024}. The trends are the same for $\delta V_s<0$, as $\ell_x^0(-\delta V_s)=\ell_y^0(\delta V_s)$, $\ell_y^0(-\delta V_s)=\ell_x^0(\delta V_s)$, $g_1^0(-\delta V_s)=-g_2^0(\delta V_s)$ and $g_2^0(-\delta V_s)=-g_1^0(\delta V_s)$.

We quantify the squeezing by the aspect ratio 
\begin{equation}
    r_\ell=\frac{\ell_y}{\ell_x}\,.
\end{equation}
In Fig. \ref{fig:sq_ideal}c we plot a map $r_\ell^0$ as a function of the confinement potential $V_C$ and the squeezing potential $V_{L,R}=-V_{T,B}=\delta V_s$. We also provide the maps of $|g_1^0|$, $|g_2^0|$ and the angle $\alpha^0$ between the precession axes of two equivalent yet orthogonally squeezed dots with $\boldsymbol{B}\parallel\boldsymbol{x}+\boldsymbol{y}$. In the blank area of the three panels the squeezed QD is not stable anymore, as it splits into strongly coupled double dots along the weak confinement direction $y$. We could hardly reach $r_\ell^0>3$ in the simulations, and only for small QDs ($\ell_c^0<18$\,nm). Remarkably, not much squeezing is required to achieve a significant contrast between $|g_1^0|$ and $|g_2^0|$, which goes through zero for moderate $r_\ell^0=1.54$ at $\ell_c^0=16$\,nm. When $g_2^0=0$, $\alpha^0$ is $90\degree$, which is the optimal angle allowing to reach any point of the Bloch sphere with a single back-and-forth hopping. Once $g_2^0$ has changed sign, we even achieve $\alpha^0>90\degree$, which does not, however, provide any significant advantage when $\boldsymbol{B}\parallel\boldsymbol{x}+\boldsymbol{y}$. The condition $\alpha^0=90\degree$ is then actually met for slightly off-plane magnetic fields. We further discuss the geometry of the optimal operation points ($\alpha=90\degree$) on the unit sphere describing the magnetic field orientation in Appendix \ref{app:optimal}. Unlike for circular QDs where the confinement does not have a strong impact on spin manipulation (see Appendix \ref{app:dep_conf}), working with small QDs is advantageous when squeezing. For large QDs the stability window is indeed significantly narrower, to the point that the optimal $g_2^0=0$ bias gets out of reach.

These results show that we can reliably achieve a finite angle between the precession axes of orthogonally squeezed QDs, which is dependent on the $g$ factor anisotropy and magnetic field orientation, and thus can be electrically engineered. This indeed makes spin manipulation by hopping disorder-independent, as $\alpha$ is finite in the pristine device. Disorder is, therefore, not used as a resource, yet it remains an open question whether it can compromise the reliability of electrical squeezing. If the effects of disorder do not supersede electrical confinement, the magnetic axes shall remain tightly locked, and the engineered $g$ factors be solid. The proposed strategy would then make spin manipulation by hopping deterministic, reliable and robust to disorder. We answer these questions in the next section.

\subsection{Spin manipulation in disordered QDs} \label{sec:sq_disorder}

\begin{figure}[t]
\centering
\includegraphics[width=0.98\columnwidth]{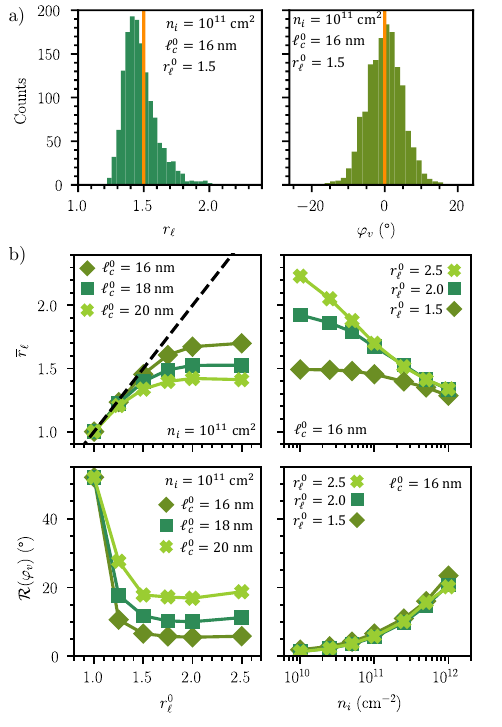}
\caption{Resilience of squeezing and axis locking to disorder. a) Histogram of the squeezing ratio $r_\ell$ and of the magnetic angle $\varphi_v$ for QDs with charge trap density $n_i=10^{11}$\,cm$^{-2}$, nominal size $\ell_c^0=16$\,nm and target squeezing ratio $r^0_\ell=1.5$ in the absence of disorder. b) Dependence of the median squeezing ratio $\overline{r}_\ell$ (top row) and of ${\cal R}(\varphi_v)$ (bottom row) on the target squeezing ratio $r_\ell^0$ (first column) and on the charge trap density $n_i$ (second column). We provide data for different nominal dot sizes $\ell_c^0=16$, $18$, $20$\,nm for the dependence on $r_\ell^0$; and data for $r_\ell^0=1.5$, $2.0$, $2.5$ for the dependence on $n_i$.}
\label{fig:sq_ax}
\end{figure}

We collect statistics on squeezed QDs with the same source of disorder as in Sec. \ref{sec:circ_dis}. In Fig. \ref{fig:sq_ax}a, we show the histogram of $r_\ell$ and $\varphi_v$ for a dataset of QDs with charge trap density $n_i=10^{11}$\,cm$^{-2}$, nominal dot size $\ell_c^0=16$\,nm, and target squeezing ratio $r_\ell^0=1.5$. We observe that the distribution of $r_\ell$'s remains centered at $\overline{r}_\ell\approx r_\ell^0\approx 1.5$, with a finite yet limited width. Remarkably, a squeezing ratio of only 1.5 is sufficient to lock the magnetic axes efficiently, as illustrated by the distribution of $\varphi_v$'s. Unlike the almost uniform distribution observed for circular QDs in Fig. \ref{fig:angles}a, the dispersion of $\varphi_v$ is only $\mathcal{R}(\varphi_v)=6.6\degree$ around $\overline{\varphi}_v=0\degree$. 

\begin{figure}[t]
\centering
\includegraphics[width=0.98\columnwidth]{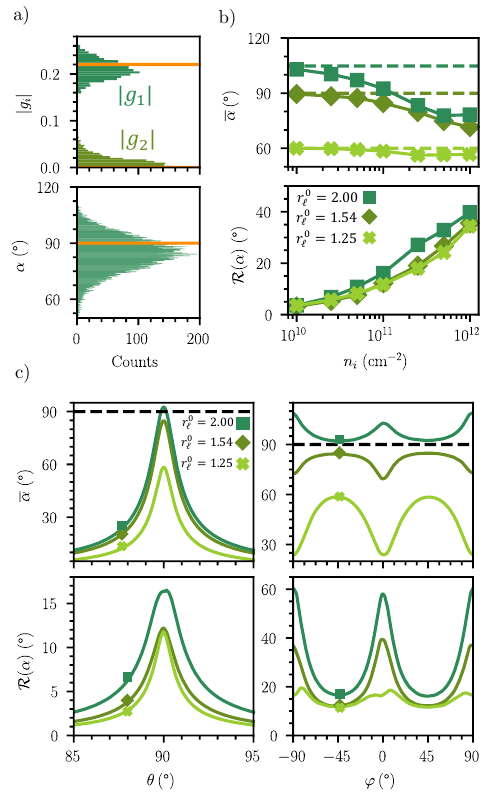}
\caption{Statistics of the angle $\alpha$ between the Larmor vectors of orthogonally squeezed QDs in the presence of disorder. a) Histograms of the principal $g$ factors $g_1$ and $g_2$, and of $\alpha$ ($\boldsymbol{B}\parallel\boldsymbol{x}+\boldsymbol{y}$) for QDs with charge trap density $n_i=10^{11}$\,cm$^{-2}$, nominal size $\ell_c^0=16$\,nm, and target squeezing ratio $r_\ell^0=1.54$. b) Dependence of $\overline{\alpha}$ and $\mathcal{R}(\alpha)$ on the charge trap density $n_i$ for $\ell_c^0=16$\,nm and $r_\ell^0=1.25,1.54,2.00$. c) Dependence of $\overline{\alpha}$ (top row) and $\mathcal{R}(\alpha)$ (bottom row) on the orientation of the magnetic field for $n_i=10^{11}$\,cm$^{-2}$, $\ell_c^0=16$\,nm and $r_\ell^0=1.25,1.54,2.00$. The left column displays the dependence on $\theta$ for $\varphi=45\degree$, and the right column the dependence on $\varphi$ for $\theta=90\degree$.}
\label{fig:sq_ang}
\end{figure}

In Fig. \ref{fig:sq_ax}b we perform a more systematic analysis of the robustness of the squeezing and axis locking to disorder. First, we discuss the dependence of $\overline{r}_\ell$ and $\mathcal{R}(\varphi_v)$ on the squeezing ratio $r_\ell^0$ at $n_i=10^{11}$\,cm$^{-2}$. We see that generally $\overline{r}_\ell<r_\ell^0$, especially for big dots (large $\ell_c^0$) and for more aggressive squeezing (large $r_\ell^0$). As a matter of fact, $\overline{r}_\ell$ remains always smaller than 1.75 for $\ell_c^0=16$\,nm, and attempting to squeeze beyond $r_\ell^0=2$ does not bring significant improvement in the presence of disorder. This is tightly related to the QD stability problem discussed in the pristine device. Large $r_\ell^0>2$ compromises the stability of the dot in the presence of disorder, and strong squeezing is then overturned by enhanced variability of the QD position. Nonetheless, $r_\ell^0\geq1.5$ is sufficient to lock the magnetic axes efficiently, especially for small QDs, which are less polarizable thus less susceptible to disorder [$\mathcal{R}(\varphi_v)<10.6 \degree$ for $r_\ell^0>1.5$ and $\ell_c^0<16$\,nm].

The dependence on $n_i$ highlights the same trends. We focus on $\ell_c^0=16$\,nm in Fig. \ref{fig:sq_ax}b, and compare different $r_\ell^0$. The top-right panel of Fig. \ref{fig:sq_ax}b shows that targeting larger $r_\ell^0$ may indeed yield larger $\overline{r}_\ell$ for very low disorder, but that all $r_\ell^0$ data merge into the same curve for large $n_i$. Indeed, electrical squeezing becomes less efficient as it gets superseded by disorder, and so does the axis locking. $\mathcal{R}(\varphi_v)$ is as large as $23\degree$ for $n_i=10^{12}$\,cm$^{-2}$, which is nonetheless still significantly better than the IQR of a uniform distribution. 

\begin{figure}[t]
\centering
\includegraphics[width=0.98\columnwidth]{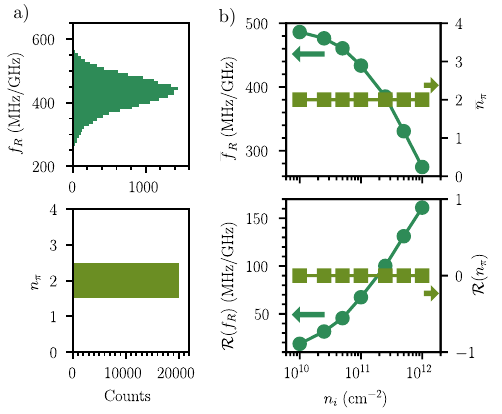}
\caption{Statistics on spin manipulation by hopping between orthogonally squeezed QDs. a) Histogram of the Rabi frequency $f_R$ and number of shuttling pulses $n_\pi$ required to perform a $\pi$ rotation for disordered QDs with charge trap density $n_i=10^{11}$\,cm$^{-2}$. b) Median (top panel) and inter-quartile range (bottom panel) of $f_R$ and $n_\pi$ as a function of $n_i$. The data are calculated for squeezed QDs with nominal size $\ell_c^0=16$\,nm, target squeezing ratio $r_\ell^0=1.54$, in a magnetic field $\boldsymbol{B}\parallel\boldsymbol{x}+\boldsymbol{y}$.}
\label{fig:sq_fR}
\end{figure}

The robust magnetic axes locking translates into large $\alpha$'s thanks to the engineered $g$ factor difference. In Fig. \ref{fig:sq_ang}a we provide the distributions of the principal $g$ factors of QD$_1$ and of the angle $\alpha$ between the precession axes of QD$_1$ and QD$_2$ for orthogonally squeezed dots with $n_i=10^{11}$\,cm$^{-2}$, $\ell_c^0=16$\,nm, and $r_\ell^0=1.54$. We choose this squeezing ratio as it achieves $g_2^0\approx 0$, thus the optimal $\alpha^0\approx 90\degree$. Disorder slightly scatters the $g$ factors, yet they remain close to the values engineered in the pristine device. This translates into systematically large $\alpha$'s, which now follow a near Gaussian distribution centered at a finite $\overline{\alpha}\approx 90\degree$. Note that such a distribution is free from the complications associated with the skewed histogram of Fig. \ref{fig:angles_pairs}a: the probability of having $\alpha\approx 0$ is negligible as long as $\overline{\alpha}/\mathcal{R}(\alpha)$ is large enough.

Conveniently, $\overline{\alpha}$ is electrically tunable as it depends on the squeezing strength. This is illustrated in Fig. \ref{fig:sq_ang}b, which plots $\overline{\alpha}$ and $\mathcal{R}(\alpha)$ as a function of $n_i$ for three squeezing ratios $r_\ell^0$. On the one hand, $\mathcal{R}(\alpha)$ is comparable to circular QDs (see Fig. \ref{fig:angles_pairs}c) and little dependent on $r_\ell^0$. On the other hand, the median $\overline{\alpha}$ slowly decreases with increasing $n_i$ as disorder tends to limit the actual squeezing of the dot (see Fig. \ref{fig:sq_ang}). It shall asymptotically approach the data for circular dots whatever $r_\ell^0$. Nevertheless, $\overline{\alpha}$ remains consistently above $\approx 60\degree$ for $n_i<10^{12}$\,cm$^{-2}$ and all $r_\ell^0$ considered here. This is actually the main advantage of orthogonally squeezed QDs: while this does not improve the variability of $\alpha$'s, this enables robust rotations ($\mathcal{R}(\alpha)\gg\overline{\alpha}$) owing to the solid imbalance between $g_1$ and $g_2$. Note that $\overline{\alpha}$ remains close to the optimal $\overline{\alpha}=90\degree$ over a wide range of trap densities for $r_\ell^0=1.54$. As for the pristine device, it can even be greater than $90\degree$ at large $r_\ell^0$, and thus reaches $90\degree$ at finite disorder strength. $r_\ell^0>1.54$ may thus be targeted for optimal operation in disordered heterostructures.


The dependence of $\overline{\alpha}$ and $\mathcal{R}(\alpha)$ on the magnetic field orientation is shown in Fig. \ref{fig:sq_ang}c. For in-plane magnetic fields ($\theta=90\degree$), $\overline{\alpha}$ is maximal at the expected $\varphi=\pm 45\degree$. The dependence on $\theta$ (for $\varphi=45\degree$) shows the same rapid out-of-plane decay as in circular QDs (Fig. \ref{fig:angles_pairs}b) due to the strong gyromagnetic anisotropy. Interestingly, for $r_\ell^0=2$ we find two optimal ($\overline{\alpha}=90\degree$) orientations for slightly off-plane magnetic fields (as expected when $g_2^0>0$). These optimal orientations actually form two parallel lines (at constant $\theta$) on the unit sphere describing the magnetic field orientation, as shown in Appendix \ref{app:optimal}.

The benefits of hopping between orthogonally squeezed QDs are best illustrated by $f_R$ and $n_\pi$, and their respective scaling with disorder. A comparison between Fig. \ref{fig:sq_fR} and Fig. \ref{fig:fR} highlights the advantages. First, the tailed distribution of $f_R$ for circular QDs becomes a Gaussian distribution for squeezed QDs. Moreover, while $\mathcal{R}(f_R)$ is roughly similar and scales the same way with $n_i$ in circular and squeezed QDs, $\overline{f}_R$ is systematically larger for squeezed QDs, and follows a trend opposite to circular QDs. Indeed, circular QDs require disorder to achieve significant angles between precession axes, while the maximal angles are reached in the absence of disorder when squeezing. Moreover, due to the large $\overline{\alpha}$, the 20\,000 squeezed QD pairs sampled within the $n_i=10^{11}$\,cm$^{-2}$ dataset all feature $n_\pi=2$, which is a major improvement with respect to the tailed distribution of Fig. \ref{fig:fR}a. In fact, the median $n_\pi$ is actually 2 and the IQR 0 for all studied $n_i\le10^{12}$\,cm$^{-2}$. While there may certainly be outlying devices requiring $n_\pi>2$, these results illustrate the remarkable robustness of the spin manipulation by hopping between squeezed QDs. Indeed, disorder is no longer used as an asset, as the medians of all relevant properties are weakly dependent of disorder. As a consequence, optimal operation falls back to requiring low disorder to minimize $\mathcal{R}$ and ensure homogeneity across the array. This requirement, same as for standard EDSR \cite{Martinez2025}, is more in line with the attended improvement of materials and interfaces quality upon scaling. 

\section{Conclusions} \label{sec:conclusions}

In this work, we have investigated the performance of hole spin manipulation by hopping in realistic Ge/GeSi devices in the presence of disorder. We have quantified the angle $\alpha$ between the precession axes of pairs of nominally circular quantum dots, and have computed the corresponding Rabi frequencies and the number of shuttling pulses required to perform a $\pi$ rotation of the spin. We find that circular QDs can indeed benefit from disorder that scatters the precession axes; however, large $\alpha$'s are only achieved for very high levels of disorder (above $n_i=5\times 10^{11}$ charge traps/cm$^2$ at the GeSi/gate stack interface). Furthermore, the resulting distribution of angles inherently includes a substantial fraction of pairs with very small $\alpha$ that require a large number of shuttling cycles to achieve spin rotations. This behavior appears incompatible with scalability roadmaps that target cleaner, more stable, and less noisy devices to improve all other performance metrics, and it therefore compromises the scaling of disorder-assisted spin manipulation by hopping.

As an alternative, we have shown that orthogonally squeezing pairs of QDs locks their magnetic axes and allows to engineer the $g$ factors so as to achieve deterministic tilts of the precession axes, yielding large and well-defined $\alpha$'s. We demonstrate that this approach effectively decouples spin manipulation from disorder, enabling reliable operation even in pristine, defect-free devices. In particular, electrostatic squeezing can be used to switch the sign of one principal in-plane $g$ factor and reach the optimal $\alpha=90^\circ$. Remarkably, both the axis locking and the $g$ factor engineering remain robust against disorder for moderate charge trap densities ($n_i\leq 5\times 10^{11}$\,cm$^{-2}$), which ensures minimal and consistent number of shuttling pulses for spin manipulation. The proposed strategy therefore enables spin manipulation by hopping in the high-quality devices required to scale up the Ge/GeSi spin qubit platform, and constitutes an appealing protocol for spin control in large-scale quantum processors.

\section*{Acknowledgments}

We thank Corentin Déprez and Esteban A. Rodriguez-Mena for fruitful discussions. This work was supported by the ``France 2030'' program (PEPR PRESQUILE-ANR-22-PETQ-0002), the French National Research Agency (project InGeQT), and by the Horizon Europe Framework Program (grant agreement 101174557 QLSI2). It was performed using HPC resources from GENCI-IDRIS (Grants 2024-AD010915504 and 2025-A0180616149). JCAU is supported by MICIU/AEI/10.13039/501100011033, and European Union Next Generation EU/PRTR through Grants RYC2022-037527-I and PID2023-148257NA-I00. JCAU also acknowledges the support of the Quantum Technologies Platform (QTEP-CSIC) and of the Severo Ochoa Centres of Excellence program through Grant CEX2024-001445-S.

\appendix

\section{Correlation effects in double quantum dots}\label{app:corr_double}

In the main text, we have assumed that the QDs are uncorrelated - namely, that the charge traps do not introduce dot-to-dot correlations between the quantities relevant for the calculation of $\alpha$, which can, therefore, be assessed from single dot calculations. In this Appendix, we validate this assumption by simulating pairs (instead of single) QDs. We show that, indeed, the spin properties of the ``left'' (QD$_1$) and ``right'' (QD$_2$) dots do not show any significant correlations. This also demonstrates that the $2\times 2$ supercell used for single dot calculations does not introduce spatial correlations due to periodic boundary conditions in the charge traps potential.

\begin{figure}[t!]
\centering
\includegraphics[width=0.98\columnwidth]{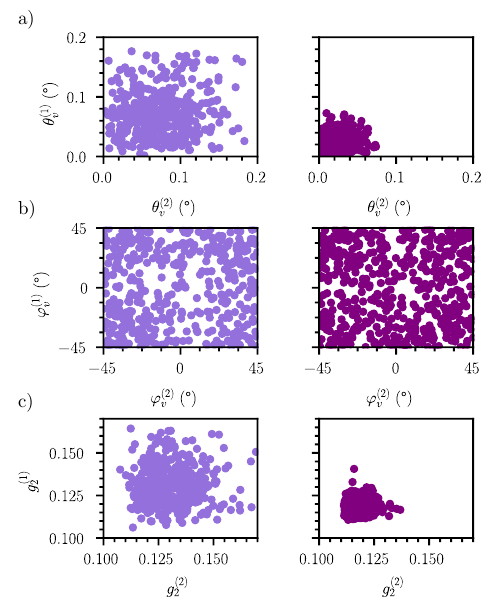}
\caption{Correlations between the properties of the left (QD$_1$) and right (QD$_2$) dots of the double QD device of Fig. \ref{fig:dev}a, for charge trap densities $n_i=10^{11}$ (left column) and $n_i=10^{10}$\,cm$^{-2}$ (right column). a), b) Dot-to-dot correlations of the magnetic angles $\theta_v$ and $\varphi_v$. c) Dot-to-dot correlations of the principal in-plane $g$ factors $g_1$ and $g_2$. For both the left and right QD simulations, the corresponding $C$ gate is biased so that the nominal dot size is $\ell_c^0=20$\, nm, with all other gates grounded.}
\label{fig:app_corr}
\end{figure}

For that purpose, we simulate the whole $3\times 2$ supercell of Fig. \ref{fig:dev}a for independent realizations of the disorder. For each realization, we bias the device in order to form the QD either under the left or right plunger, and we compute the magnetic angles $\theta_v$ and $\varphi_v$, as well as the principal $g$ factors $g_1$ and $g_2$ in each dot. We plot the correlations between the dots in Fig. \ref{fig:app_corr}, for $n_i=10^{10}$ and $n_i=10^{11}$\,cm$^{-2}$. None of the above quantities show clear dot-to-dot correlations, even at low trap densities. As a consequence, the left and right QDs can be assumed independent, and the properties of double QDs computed by pairing single dot simulations.

\section{Spin basis alignment}\label{app:spin_align}

We compute the ground-state orbitals $\ket{\Uparrow}$ and $\ket{\Downarrow}$ of single QDs using a finite-difference implementation of the Luttinger-Kohn Hamiltonian. We emphasize that the pseudo-spins $\Uparrow$ and $\Downarrow$ are defined up to a unitary transform as the ground-state is degenerate (Kramers degeneracy due to time-reversal symmetry). We then consider double QDs built on pairs (QD$_1$, QD$_2$) of single QDs with ground doublets $\{\ket{1\Uparrow},\ket{1\Downarrow}\}$ and $\{\ket{2\Uparrow},\ket{2\Downarrow}\}$. We shuttle a hole between the two dots every half Larmor period in order to drive coherent rotations of the spin \cite{wang2024}. We assume that the tunnel coupling and control pulses are tuned so that the hole is shuttled adiabatically with respect to the charge degree of freedom but diabatically with respect to the spin degree of freedom. An initial state $\ket{\varphi^{(1)}}=\alpha\ket{1\Uparrow}+\beta\ket{1\Downarrow}$ in QD$_1$ thus shuttles to a final state $\ket{\varphi^{(2)}}$ in the $\{\ket{2\Uparrow},\ket{2\Downarrow}\}$ subspace of QD$_2$. To work out this final state, we first transform the orbitals $\{\ket{2\Uparrow},\ket{2\Downarrow}\}$ of QD$_2$ into orbitals $\{\ket{2\Uparrow'},\ket{2\Downarrow'}\}$ such that $\mathrm{Re}(\braket{2\Uparrow'|1\Uparrow})$ and $\mathrm{Re}(\braket{2\Downarrow'|1\Downarrow})$ are maximum, and $\braket{2\Uparrow'|1\Downarrow}=\braket{2\Downarrow'|1\Uparrow}=0$. We then assume that $\ket{1\Uparrow}$ shuttles to $\ket{2\Uparrow'}$ and $\ket{1\Downarrow}$ shuttles to $\ket{2\Downarrow'}$ (and vice-versa). Therefore,
\begin{equation}
\ket{\varphi^{(2)}}=\alpha\ket{2\Uparrow'}+\beta\ket{2\Downarrow'}\,.
\end{equation}
The above conditions are actually met by the transformation $\ket{2\Uparrow'}=T_{11}\ket{2\Uparrow}+T_{12}\ket{2\Downarrow}$ and $\ket{2\Downarrow'}=T_{21}\ket{2\Uparrow}+T_{22}\ket{2\Downarrow}$, where \cite{Venitucci2018}
\begin{equation}
T=\beta\begin{pmatrix}
  \braket{2\Uparrow|1\Uparrow}   & \braket{2\Downarrow|1\Uparrow} \\
  \braket{2\Uparrow|1\Downarrow} & \braket{2\Downarrow|1\Downarrow}
  \end{pmatrix}
\end{equation}
and
\begin{subequations}
\begin{align}
\beta^{-2}&=|\braket{2\Uparrow|1\Uparrow}|^2+|\braket{2\Downarrow|1\Uparrow}|^2 \\
          &=|\braket{2\Uparrow|1\Downarrow}|^2+|\braket{2\Downarrow|1\Downarrow}|^2\,.
\end{align}
\end{subequations}
The transformation $T$ is always unitary if $\{\ket{1\Uparrow},\ket{1\Downarrow}\}$ and $\{\ket{2\Uparrow},\ket{2\Downarrow}\}$ are both Kramers pairs. 

With these assumptions, the angle $\alpha$ between the precession axes of QD$_1$ and QD$_2$ can be calculated with Eqs.~\eqref{eq:fL} and ~\eqref{eq:alpha} using the $g$ matrices $\hat{g}^{(1)}$ in the $\{\ket{1\Uparrow},\ket{1\Downarrow}\}$ basis set and $\hat{g}^{(2)}$ in the ``aligned'' $\{\ket{2\Uparrow'},\ket{2\Downarrow'}\}$ basis set. Note that aligning $\{\ket{1\Uparrow'},\ket{1\Downarrow'}\}$ on $\{\ket{2\Uparrow},\ket{2\Downarrow}\}$ (with the matrix $T^\dagger$) yields the same $\alpha$ (as it should).

In disordered devices, we draw QD$_1$ and QD$_2$ from a set of $N$ random realizations of disorder with ground doublets $\{\ket{R_n\Uparrow},\ket{R_n\Downarrow}\}$ ($1\le n\le N$). Instead of aligning $\{\ket{2\Uparrow},\ket{2\Downarrow}\}$ on $\{\ket{1\Uparrow},\ket{1\Downarrow}\}$ for each pair, we directly pick $\hat{g}^{(1)}$ and $\hat{g}^{(2)}$ in the set $\{\hat{g}_n\}$ of $g$ matrices computed in the basis sets $\{\ket{R_n\Uparrow'},\ket{R_n\Downarrow'}\}$ aligned on the pristine doublet $\{\ket{P\Uparrow},\ket{P\Downarrow}\}$. We thus only need to store the $\hat{g}_n$'s, and not the wave functions $\{\ket{R_n\Uparrow},\ket{R_n\Downarrow}\}$ of all realizations (which represent terabytes of data). Although this procedure is not strictly equivalent to the alignment of $\{\ket{2\Uparrow},\ket{2\Downarrow}\}$ on $\{\ket{1\Uparrow},\ket{1\Downarrow}\}$ (it is only exact to first-order in the disorder strength), tests on subsets of realizations show that the error on $\alpha$ is typically $\lesssim 0.5\degree$.

As discussed in the main text, the pristine basis set $\{\ket{P\Uparrow},\ket{P\Downarrow}\}$ can always be chosen so that $\hat{g}^0$ is diagonal ($\hat{U}^0=\hat{V}^0=\hat{I}$ when the magnetic field is expressed in the $\{\boldsymbol{x},\boldsymbol{y},\boldsymbol{z}\}$ frame). The choice of $\{\ket{P\Uparrow},\ket{P\Downarrow}\}$ has, nevertheless, no impact on the $\alpha$'s (that are observables). It only shifts the Larmor angles $\theta_L$ and $\varphi_L$ (since the spin axes $\hat{U}^0$ may not match $\{\boldsymbol{x},\boldsymbol{y},\boldsymbol{z}\}$ for arbitrary $\{\ket{P\Uparrow},\ket{P\Downarrow}\}$). We do not, however, work out the spin basis set that diagonalizes $\hat{g}^0$ explicitly; instead, we compute $\hat{U}^0$ for the raw doublet returned by the eigensolver, and define the Larmor vector of all dots (whether pristine or disordered) as
\begin{equation}
   h\boldsymbol{f}_L=\mu_B(\hat{U}^0)^\intercal\hat{g}_n\boldsymbol{B}\,,
\end{equation}
which expresses $\boldsymbol{f}_L$ in the spin axes of the pristine device that we map onto $\{\boldsymbol{x},\boldsymbol{y},\boldsymbol{z}\}$.

\section{Impact of detuning on the precession axes} \label{app:doubleDot}

\begin{figure}[t!]
\centering
\includegraphics[width=0.98\columnwidth]{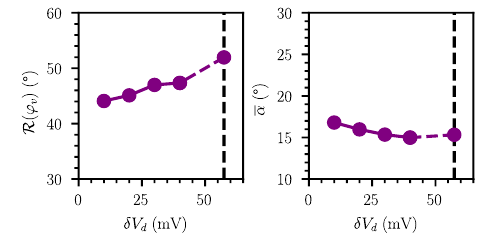}
\caption{Dependence of $\mathcal{R}(\varphi_v)$ and $\overline{\alpha}$ on the detuning $\pm\delta V_d$ on both ends of the hopping sequence for double QDs with charge trap density $n_i=10^{11}$\,cm$^{-2}$. The dashed vertical line highlights the nominally circular single QD limit ($\ell_c^0=20$\,nm) discussed in the main text.}
\label{fig:app_depDet}
\end{figure}

In Section \ref{sec:circ} we have assumed that the hole is shuttled between ``fully symmetric'' bias configurations where the dot is nominally circular (all gates grounded except the plunger gate $C_1$ of QD$_1$ or $C_2$ of QD$_2$). On both ends of the shuttling sequence (hole in QD$_1$ or QD$_2$), the average plunger gate voltage is thus $\overline{V}_C=(V_{C_1}+V_{C_2})/2=V_C/2$ (set to reach a target dot size $\ell_c^0$), while the detuning voltage is $\delta V_d=V_{C_2}-V_{C_1}=\pm V_C$. However, the dots may practically be shuttled between different detunings, where the wave functions of both dots are slightly squeezed perpendicular to the tunneling axis by the imbalance between $V_{C_1}$, $V_{C_2}$ and the surrounding barrier and plunger gate voltages \footnote{We emphasize that both dots are then squeezed along the same axis, contrary to section \ref{sec:squeezed}.}. Here we show that the resulting corrections are small, and do not significantly alter the conclusions presented in the main text.

For that purpose, we simulate a double QD device in the $3\times 2$ supercell of Fig. \ref{fig:dev}a. We consider nominally circular dots with size $\ell_c^0=20$\,nm in a charge trap density $n_i=10^{11}$\,cm$^{-2}$ ($V_C=-57.4$\,mV and $\delta V_d=\pm V_C$). To assess the impact of detuning on axis locking and precession, we track $\mathcal{R}(\varphi_v)$ and $\overline{\alpha}$ as a function $|\delta V_d|<|V_C|$ for fixed $V_C=-57.4$\,mV. As shown in Fig. \ref{fig:app_depDet}, the dependence of both properties on $\delta V_d$ is rather weak: $\mathcal{R}(\varphi_v)$ is slightly reduced by the detuning-induced symmetry breaking and axis locking, while $\overline{\alpha}$ is enhanced by the corresponding $g$ factor variations. Additionally, operating closer to the charge anti-crossing (small $|\delta V_d|$) may give rise to extra renormalizations of the principal $g$ factors due to magneto-tunneling terms \cite{Rodriguez2025}. Yet the results at detuning $|\delta V_d|<57.4$\,mV are qualitatively, and to a large extent quantitatively similar to those of the main text, especially at large trap densities where disorder supersedes the effects of detuning.

\section{Disorder-induced fluctuations of the spin axes}\label{app:corr_spin}

In the main text, we have identified three mechanisms that scatter the precession axes: the fluctuations of the spin axes $\hat{U}$, of the magnetic axes $\hat{V}$, and of the principal $g$ factors $g_1$, $g_2$, $g_3$. We have provided data for the latter two in Figs. \ref{fig:angles}a and \ref{fig:angles_pairs}a. In this Appendix, we quantify the fluctuations of the spin axes. 

Similarly to the magnetic angles $\theta_v$ and $\varphi_v$, we introduce two angles $\theta_u$ and $\varphi_u$ that characterize the fluctuations of the spin axes. We define $\theta_u$ as the polar angle between $\boldsymbol{u_3}$ and $\boldsymbol{u}_3^0\equiv\boldsymbol{z}$, and $\varphi_u$ as the azimuthal angle between the projection of $\boldsymbol{u}_1$ in the $xy$ plane and $\boldsymbol{u}_1^0\equiv\boldsymbol{x}$. 

\begin{figure}[t!]
\centering
\includegraphics[width=0.98\columnwidth]{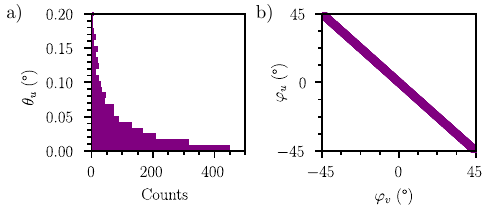}
\caption{Disorder-induced scattering of the spin axes. a) Histogram of the angle $\theta_u$ between the spin axis $\boldsymbol{u}_3$ and $\boldsymbol{u}_3^0\equiv\boldsymbol{z}$ for circular QDs with $\ell_c^0=20$\,nm and $n_i=10^{11}$\,cm$^{-2}$. b) Correlations between $\varphi_u$ and $\varphi_v$ for the same dataset as in a).}
\label{fig:app_spin}
\end{figure}

In Fig. \ref{fig:app_spin}a we plot the histogram of $\theta_u$ for the same dataset of circular QDs as in Fig. \ref{fig:angles}a ($n_i=10^{11}$\,cm$^{-2}$). The $\theta_u$'s follow a tailed distribution peaked at $0\degree$, and are generally smaller than $\theta_v$. There are, moreover, no clear correlations between $\theta_u$ and $\theta_v$. The finite $g_{31}$ and $g_{32}$ resulting from the coupling between the in- and out-of-plane motions as well as the shear strains $\varepsilon_{xz}$ and $\varepsilon_{yz}$ indeed tilt the magnetic axes more than the spin axes (as $g_{13}\approx g_{23}\approx 0$) \cite{Martinez2022-inhom,Abadillo2022,Mauro2025}. 

Interestingly, $\varphi_u$ perfectly correlates with $-\varphi_v$. As a matter of fact, perturbation theory predicts that $g_{12}=-g_{21}$ (these elements being non-zero due to the coupling between the motions along $x$ and $y$ and due to the shear strains $\varepsilon_{xy}$) \cite{Michal2021,Abadillo2022}. Therefore, any in-plane rotation of $\boldsymbol{v}_1$, $\boldsymbol{v}_2$ gives rise to an opposite rotation of $\boldsymbol{u}_1$, $\boldsymbol{u}_2$. 

As $\theta_u\ll\theta_v$ and $\varphi_u\approx-\varphi_v$, the orientation of the precession axes of the QDs can be understood in the present germanium heterostructures from the analysis of the principal magnetic axes and $g$ factors, as done in the main text.

\section{Proposal for a scalable device architecture} \label{app:sq_architecture}

\begin{figure}[t!]
\centering
\includegraphics[width=0.98\columnwidth]{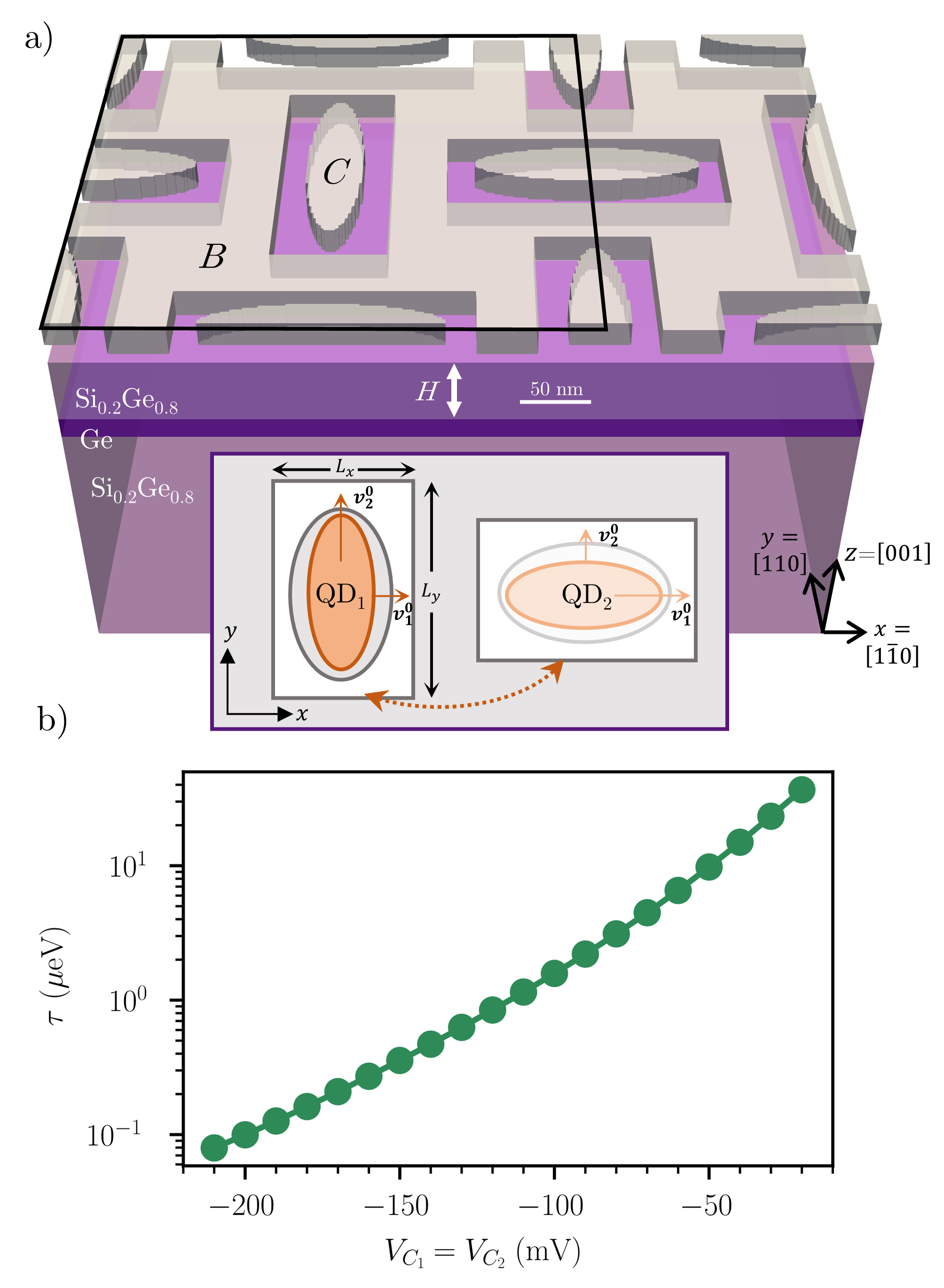}
\caption{a) Proposed device architecture with reduced overhead. A two-dimensional array of elliptical plunger gates $C$, rotated by $90\degree$ from one site to the other, is surrounded by a common, shared $B$ gate. The inset illustrates how $L_x\neq L_y$ squeezes the QDs, which locks the magnetic axes $\boldsymbol{v}_1$, $\boldsymbol{v}_2$ and shapes the principal $g$ factors $g_1$, $g_2$. b) Dependence of the tunnel coupling $\tau$ between two neighboring dots on the confinement strength, modulated by the plunger gates voltages $V_{C_1}=V_{C_2}$ ($V_B=0$\,V, $L_x=80$\,nm, $L_y=200$\,nm, and $H=50$\,nm).}
\label{fig:app_layoutDev}
\end{figure}

In this Appendix, we propose a scalable gate layout with minimal overhead that ensures sufficient squeezing for deterministic spin manipulation by hopping. The device is illustrated in Fig. \ref{fig:app_layoutDev}a. It comprises a two-dimensional array of plunger gates $C$ intertwined with a single barrier gate $B$ shared by the whole array. The dots are geometrically squeezed by the asymmetric shape of the $C$ gates and $B$ gate around. These shapes are rotated by $90\degree$ from one dot to the other. Such an array requires $N+1$ control knobs for $N$ dots, an improvement with respect to the $3N$ knobs needed by the device of Fig. \ref{fig:dev}a. 

Nevertheless, this array does not have dedicated gates for each inter-dot barrier. It shall thus be sparsely occupied to enable spin manipulation by hopping and selective exchange interactions. The tunnel coupling and hopping between neighboring sites are controlled by the $C$ gate voltages, as already demonstrated in a similar layout with circular QDs \cite{ivlev2025}. We illustrate the controllability of the tunnel coupling $\tau$ in a double QD in Fig. \ref{fig:app_layoutDev}b. Remarkably, we are able to tune $\tau$ over two orders of magnitude in a 150\,mV window (for a device with top SiGe barrier thickness $H=50$\,nm).

\begin{figure}[t!]
\centering
\includegraphics[width=0.98\columnwidth]{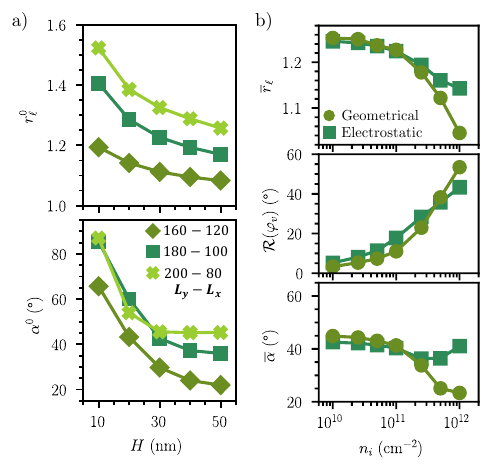}
\caption{a) Dependence of $r_\ell^0$ and $\alpha^0$ on the top GeSi barrier thickness $H$ for the device of Fig. \ref{fig:app_layoutDev}a with different ratios between $L_x$ and $L_y$. We set $V_B=0$ and $V_{C}=-25.63$, $-36.40$, $-52.78$, $-74.63$, $-102.60$\,mV for $H=10$, $20$, $30$, $40$, $50$\,nm, respectively, so that $\ell_c^0=16$\,nm. b) Dependence of $\overline{r_\ell}$, $\mathcal{R}(\varphi_v)$ and $\overline{\alpha}$ on the charge trap density $n_i$ for geometrical squeezing ($H=50$\,nm $L_x=80$\,nm, $L_y=200$\,nm) and electrostatic squeezing (for equivalent $\ell_x^0$, $\ell_y^0$).}
\label{fig:app_layoutData}
\end{figure}

Squeezing in the array is controlled by the lengths $L_x$ and $L_y$ of the rectangular openings in the B gate (see Fig. \ref{fig:app_layoutDev}a). The gap between the elliptical $C$ gates and the $B$ gate is at least $20$\,nm. This gate pattern indeed imprints an anisotropic confinement potential in the heterostructure below. We provide the dependence of the squeezing ratio $r_\ell^0=\ell_y^0/\ell_x^0$ on the top GeSi barrier thickness $H$ for different $L_x$ and $L_y$ in Fig. \ref{fig:app_layoutData}a. We observe that the squeezing ratio remains significantly smaller than $L_y/L_x$, and steadily decreases with increasing $H$. As a matter of fact, the strongly patterned potential at the GeSi/Al$_2$O$_3$ interface softens when going deeper in the semiconductor. For $H=50$\,nm, we reach $r_\ell^0\approx 1.25$ for the most aggressive gate layout ($L_x=80$\,nm, $L_y=200$\,nm). Thinner  GeSi barriers actually make squeezing more efficient, $r_\ell^0$ reaching up to $\gtrsim 1.5$ for $H=10$\,nm. Nevertheless, reducing $H$ also brings disorder closer to the active layer, which is expected to enhance variability. 

Remarkably, geometrical squeezing is sufficient to induce significant differences between $g_1^0$ and $g_2^0$ and achieve large $\alpha^0$ \footnote{To make use of Eq.~\eqref{eq:alpha}, we compute the $g$ matrix $\hat{g}^{(1)}$ in a $2\times 2$ supercell S$_1$ with the central dot QD$_1$ elongated along $y$ (solid line in Fig. \ref{fig:app_layoutDev}), and $\hat{g}^{(2)}$ in a $2\times 2$ supercell S$_2$ with the central dot QD$_2$ elongated along $x$.}. Even though we can hardly reach the optimal $\alpha^0=90\degree$, we achieve $n_\pi^0=2$ for $H=30$\,nm in all considered layouts. For $H=50$\,nm, $\alpha^0$ is greater than $45\degree$ only for $L_x=80$\,nm, $L_y=200$\,nm. While being less efficient than the electrostatic squeezing discussed in the main text, geometrical squeezing still enables spin manipulation by hopping in the absence of disorder, and thus presents a clear advantage with respect to hopping between nominally circular QDs.

In Fig. \ref{fig:app_layoutData}b we illustrate the resilience of the above performances to disorder. We focus on $H=50$\,nm and the strongest geometrical squeezing ($L_x=80$\,nm, $L_y=200$\,nm), which yields $r_\ell^0=1.25$ in the pristine device. We compare these data to the equivalent electrostatic squeezing (with similar $\ell_x^0$ and $\ell_y^0$ at low trap density). We observe that both the squeezing and the axis locking are less tolerant to disorder for geometrical than for electrostatic squeezing, especially at large trap density $n_i>2.5\times 10^{11}$\,cm$^{-2}$. The potential landscapes are indeed slightly different (the potential for geometrical squeezing being for example shallower between the dots). Nevertheless, both approaches provide similar results for moderate disorder, and the proposed gate layout achieves a significantly large, deterministic $\overline{\alpha}\approx 45\degree$ that remains robust to disorder up to $n_i=10^{11}$\,cm$^{-2}$. This represents a notable improvement over hopping in an array of homogeneous QDs as proposed in Ref. \cite{ivlev2025}.

\section{Spin manipulation by diabatic intra-dot squeezing}\label{app:intradot}

\begin{figure}[t!]
\centering
\includegraphics[width=0.98\columnwidth]{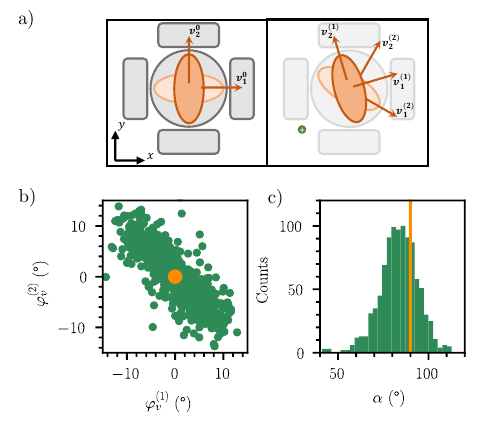}
\caption{Principles and properties of intra-dot squeezing. a) Schematic representation of a pristine device (left panel) and a device containing one charge trap (in green) with an illustrative envelope of the hole density in the QD squeezed along $x$ (dark orange) and $y$ (light orange). b) Correlations between the in-plane tilt of the magnetic axes for the state squeezed along $x$ ($\varphi_v^{(1)}$) and for the state squeezed along $y$ ($\varphi_v^{(2)}$) within the same dot, for $n_i=10^{11}$\,cm$^{-2}$, $\ell_c^0=16$\,nm and $r_\ell^0=1.54$. The orange point is the pristine device $\varphi_v^{(1)}=\varphi_v^{(2)}=0$. c) Distribution of $\alpha$ for the same dataset as in b). The orange line highlights the angle $\alpha^0=90\degree$ in the pristine device. } 
\label{fig:app_intra}
\end{figure}

\begin{figure}[t!]
\centering
\includegraphics[width=0.98\columnwidth]{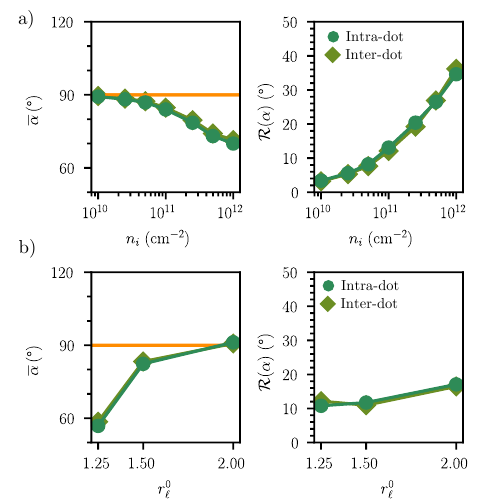}
\caption{Comparison between intra-dot and inter-dot squeezing. a) Dependence of $\overline{\alpha}$ and $\mathcal{R}(\alpha)$ on the charge trap density $n_i$ for orthogonally squeezed states within the same QD (intra-dot) and for hopping between orthogonally squeezed QDs (inter-dot), with $\ell^0_c=16$\,nm and $r_\ell^0=1.54$. b) Dependence of $\overline{\alpha}$ and $\mathcal{R}(\alpha)$ on the squeezing ratio $r_\ell^0$, for $n_i=10^{11}$\,cm$^{-2}$. The orange lines highlight the angle $\alpha^0=90\degree$ in the pristine device.} 
\label{fig:app_intra2}
\end{figure}

As briefly mentioned in the main text, spin manipulation by hopping between orthogonally squeezed QDs can be substituted by a sequence of orthogonal squeezings within a single QD (see Fig. \ref{fig:app_intra}a). Indeed, there is no fundamental need to move the spin between different QDs, and the shuttling pulses could be replaced by intra-dot pulses that alternatively squeeze the dot along $x$ and $y$ [every $\Delta t=1/(2f_L)$]. As long as the pulses are diabatic with respect to the spin, the two approaches are essentially equivalent for a pristine device, and shall provide the same performances. However, in disordered devices, the two orthogonally squeezed states are exposed to the same charge traps if manipulation is performed within the same QD, which can give rise to correlations. In this Appendix we show that these correlations exist, but do not statistically impact the angle $\alpha$ between the precession axes.  

To assess the correlations between orthogonally squeezed states within the same QD, we collect statistics on several hundreds of disorder realizations. We plot in Fig. \ref{fig:app_intra}b the $\varphi_v$'s of the $y$-squeezed ($\varphi_v^{(2)}$) versus the $\varphi_v$'s of the $x$-squeezed ($\varphi_v^{(1)}$) states of the same QD, for $n_i=10^{11}$ cm$^{-2}$, $\ell_c^0=16$ nm and the ``optimal'' $r_\ell^0=1.54$ ($r_\ell^0=1/1.54$ for the orthogonally squeezed state) that achieves $\alpha^0=90\degree$. These angles are the most (anti-)correlated quantities found in this dataset. As illustrated in Fig. \ref{fig:app_intra}a, the magnetic axes are indeed expected to display opposite tilts when the two states are exposed to the same disorder. Interestingly, these correlations do not have much statistical relevance, as the distribution of $\alpha$'s (Fig. \ref{fig:app_intra}c) and the dependence of $\overline{\alpha}$ and $\mathcal{R}(\alpha)$ on $n_i$ (Fig. \ref{fig:app_intra2}a) are almost equivalent for intra-dot squeezing and inter-dot hopping between squeezed QDs. 

Analytical modeling with test $g$ matrices (Gaussian $\varphi_v^{(2)}=-\varphi_v^{(1)}$, $\varphi_u=-\varphi_v$) show that these correlations have hardly any effect on $\overline{\alpha}$ but are expected to increase $\mathcal{R}(\alpha)$ by a factor $\approx\sqrt{2}$ at low trap density $n_i$. However, this trend is almost completely washed out by the weakly correlated variations of the principal $g$-factors $g_1$ and $g_2$ of the $x$- and $y$-squeezed dots, and by those of the angles $\theta_v^{(1)}$ and $\theta_v^{(2)}$. Moreover, the variability of all quantities increases but the correlations decrease with increasing $n_i$. Therefore, the first and second order statistics of $\alpha$ end up very similar for intra- and inter-dot squeezings at all trap densities from $n_i=10^{10}$ to $n_i=10^{12}$\,cm$^{-2}$. These conclusions also hold for different $r_\ell^0$'s, as shown in Fig. \ref{fig:app_intra2}b. Intra-dot dynamical squeezing is, therefore, expected to show the same performances as hopping between orthogonally squeezed QDs, but may require to act on more gates simultaneously (and thus increase power consumption).

\section{Squeezing in a double dot device}\label{app:tc}

The methodology of the main text assumes that the two QDs between which the hole is tunneling can be treated as independent units (thus neglecting crosstalk effects). Moreover, fast adiabatic charge transitions require large tunnel coupling, a condition not imposed in the main text. In this Appendix, we demonstrate that we can actually engineer orthogonal squeezing in the double quantum dot device of Fig.~\ref{fig:sq_ideal}a (through gates $L$, $R$), while leaving the central $B$ gate as control knob for the tunnel coupling.

In the main text, we have considered single dots squeezed by potentials $V_{L,R}=-V_{T,B}=\delta V_s$. In a double dot device based on the same gate layout (Fig.~\ref{fig:sq_ideal}a), we can squeeze the dots by applying $2\delta V_s$ on the top $L$, $R$ gates and $-2\delta V_s$ on the bottom $L$, $R$ gates, along with a shift $\delta V_s$ on gate $C_1$ and $-\delta V_s$ on gate $C_2$. Indeed, these single and double dot configurations are equivalent up to nearest neighbor gates (namely, the difference of potential between the C gates and the surrounding barrier gates is the same). The squeezing ratios may, however, be different due to the effect of farther gates that are biased differently.

\begin{figure}[t!]
\centering
\includegraphics[width=0.98\columnwidth]{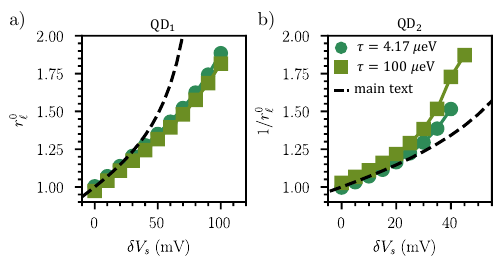}
\caption{Squeezing in a double quantum dot device. a) Dependence of $r_\ell^0$ on $\delta V_s$ in the QD squeezed along the tunneling axis, for $V_B=0$ and $V_B=-17.53$\,mV ($\tau=4.17$ and $\tau=100$\,$\mu$eV, respectively, when $\delta V_s=0$). The dashed black line is the squeezing achieved in the single dot limit of the main text. b) Dependence of $r_\ell^0$ on $\delta V_s$ for the QD squeezed perpendicularly to the tunneling axis.}
\label{fig:sq_dqd}
\end{figure}

We plot in Fig. \ref{fig:sq_dqd} the squeezing ratios computed in the double dot as a function of $\delta V_s$ for $V_B=0$\,mV (with a uniform density of charges $n_i=10^{11}$\,cm$^{-2}$ and $V_C=-117.44$\,mV in both dots so that $\ell_x^0=\ell_y^0=\ell_c^0=16$\,nm when $\delta V_s\to 0$). They are indeed close to the ratios computed in the single dot unit cell. There are, nevertheless, some noticeable differences. In particular, squeezing along the tunneling axis is harder in the double dot than in the single dot device. Moreover, squeezing perpendicular to that tunneling axis is limited to $r_\ell^0\approx 1.5$ because the hole ultimately escapes the QD. This could in principle be alleviated by screening the double dot surroundings with the outer gates (see device in Fig. \ref{fig:dev}a), grounded at present. We can nonetheless find distinct $\delta V_s$ for each QD achieving equivalent and sufficiently large $r_\ell^0$. This demonstrates that the squeezing can effectively be controlled with only a pair of side gates, leaving the central $B$ gate available for the control of the tunnel coupling $\tau$.

The tunnel coupling in the circular double dot system ($\delta V_s=0$) in the bias conditions equivalent to the main text ($V_B=0$\,mV on the central B gate and $\ell_\parallel^c=16$\,nm) is $\tau=4.17$\,$\mu$eV. By setting $V_B=-17.53$\,mV we reach $\tau=100$\,$\mu$eV, more in line with experimental operation points \cite{wang2024}. In Fig. \ref{fig:sq_dqd} we also plot the dependence of the squeezing ratio on $\delta V_s$ for $V_B=-17.53$\,mV. Remarkably, varying $V_B$ only slightly shifts the trends and we can still reach the same range of $r_\ell^0$. This shows that squeezing and tunnel coupling can be controlled rather independently in this device, thus fulfilling the requirements for robust manipulation by spin shuttling between orthogonally squeezed QDs.

The assumption of orthogonally squeezed QDs with similar squeezing ratios is therefore a reasonable starting point for section \ref{sec:squeezed}. Most of the conclusions can actually be generalized to arbitrary (different) squeezing ratios (see, e.g., Appendix \ref{app:optimal}). Moreover, this configuration can also be achieved by gate layout engineering, as discussed in Appendix \ref{app:sq_architecture}.

\section{Optimal magnetic field orientations for hopping between arbitrarily squeezed dots} \label{app:optimal}

\begin{figure}[t!]
\centering
\includegraphics[width=0.98\columnwidth]{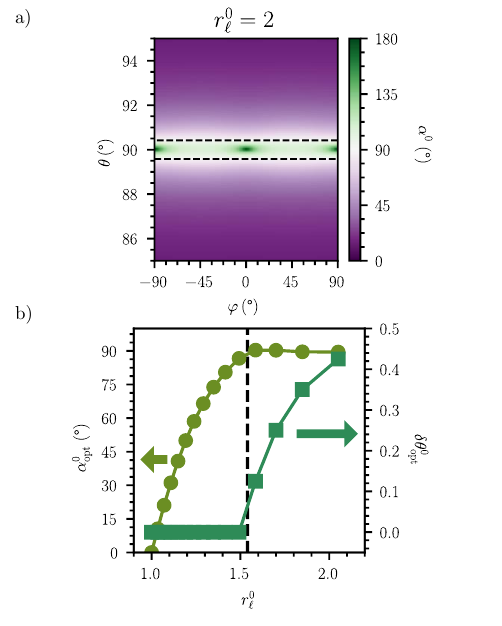}
\caption{Geometry of the optimal magnetic field orientations for a pair of orthogonally squeezed QDs in the device of Fig. \ref{fig:dev}a ($r_\ell^{(1)}=1/r_\ell^{(2)}=r_\ell^0=2$ and $\ell_c^0=16$\,nm). a) Dependence of $\alpha^0$ on the magnetic field angles $\theta$ and $\varphi$. The black dashed lines are the optimal operation lines $\alpha^0=90\degree$. b) Optimal magnetic field elevation $\delta\theta^0_{\rm opt}$, and optimal $\alpha^0_{\rm opt}$ at $\theta=90\degree\pm\delta\theta^0_{\rm opt}$, as a function of $r_\ell^0$. Note that the optimal magnetic field azimuths are $\varphi=\pm45\degree$ whatever $r_\ell^0$ if $r_\ell^{(1)}=1/r_\ell^{(2)}$. The black dashed line marks $r_\ell^0=1.54$ where $g_2^0=0$.}
\label{fig:app_optimal2}
\end{figure}

\begin{figure}[t!]
\centering
\includegraphics[width=0.98\columnwidth]{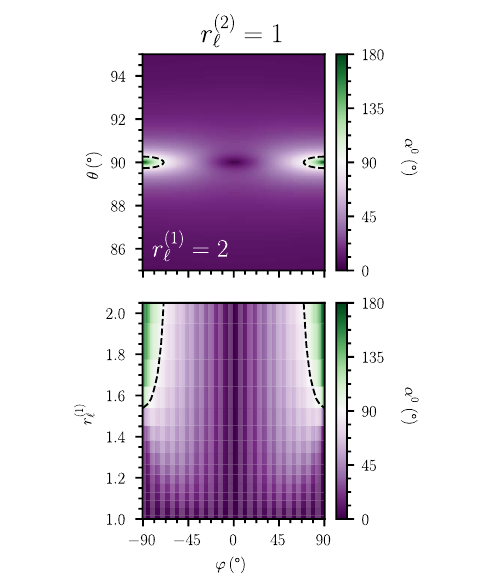}
\caption{Geometry of the optimal magnetic field orientations for hopping between an electrostatically squeezed ($r_\ell^{(1)}=2$) and a circular ($r_\ell^{(2)}=1$) QD with $\ell_c^0=16$\,nm. a) Dependence of $\alpha^0$ on the magnetic field angles $\theta$ and $\varphi$. The black dashed lines are the optimal operation lines $\alpha^0=90\degree$. b) Map of $\alpha^0$ as a function of $\varphi$ and $r_{\ell}^{(1)}$, for fixed $\theta=90\degree$. The black dashed lines highlight the optimal operation points $\alpha^0=90\degree$.} 
\label{fig:app_optimal}
\end{figure}

The discussion about orthogonally squeezed QDs in the main text only dealt with the prototypical case of hopping between a pair of dots with equivalent squeezing ratios $r_{\ell}^{(1)}=1/r_{\ell}^{(2)}=r_\ell^0$. We have highlighted on Fig. \ref{fig:sq_ang}c that the optimal magnetic field orientations lie in-plane ($\boldsymbol{B}\parallel\boldsymbol{x}\pm\boldsymbol{y}$) as long as $g_2^0\leq0$, and slightly shift out-of-plane when $g_2^0>0$. In this Appendix we generalize this discussion to pairs of QDs with arbitrary $r_{\ell}^{(1)}$, $r_{\ell}^{(2)}$, and show that we can actually engineer the optimal orientations by tuning the squeezing of each dot separately. 

As shown by Eqs.~\eqref{eq:fL} and \eqref{eq:alpha}, the angle $\alpha$ between the Larmor vectors of two QDs can be expressed as a function of the (non-symmetric) $g$ matrices of these dots. We may alternatively recast $\alpha$ in terms of symmetric Zeeman tensors \cite{Mauro2024}. The Zeeman tensor of QD$_i$ is defined as $\hat{G}^{(i)}=\hat{g}^{(i)\intercal}\hat{g}^{(i)}$; its eigenvalues are the squares of the principal $g$ factors and its eigenvectors are the principal magnetic axes of the dot ($\hat{G}=\hat{V}\hat{g}_d^2\hat{V}^\intercal$). We can additionally introduce the combined symmetric Zeeman tensor 
\begin{equation}
\hat{G}^{(12)}=\frac{1}{2}[\hat{g}^{(1)\intercal}\hat{g}^{(2)}+\hat{g}^{(2)\intercal}\hat{g}^{(1)}]\,.
\end{equation}
With these definitions we can conveniently rewrite $\cos\alpha$ as
\begin{equation}\label{eq:alpha-tensorial}
    \cos\alpha=\frac{\boldsymbol{B}\cdot\hat{G}^{(12)}\boldsymbol{B}}{\sqrt{\boldsymbol{B}\cdot\hat{G}^{(1)}\boldsymbol{B}}\sqrt{\boldsymbol{B}\cdot \hat{G}^{(2)}\boldsymbol{B}}}\,.
\end{equation}
We easily recover Eq.~\eqref{eq:alpha_sq} for two QDs with identical yet orthogonal squeezing [$\hat{g}^{(1)}={\rm diag}(g_1,g_2,g_3)$, $\hat{g}^{(2)}={\rm diag}(-g_2,-g_1,g_3)$] and an in-plane magnetic field ($\theta=90\degree$). To go further, let us consider the $g$ matrices of two QDs with arbitrary squeezings (but same magnetic axes), $\hat{g}^{(1)}={\rm diag}(g_1^{(1)},g_2^{(1)},g_3^{(1)})$, and $\hat{g}^{(2)}={\rm diag}(g_1^{(2)},g_2^{(2)},g_3^{(2)})$. In this more general case, $G^{(12)}={\rm diag}(g_1^{(1)}g_1^{(2)},g_2^{(1)}g_2^{(2)},g_3^{(1)}g_3^{(2)})$, and the optimal magnetic field orientations that achieve $\alpha=90\degree$ ($\cos\alpha=0$) fulfill
\begin{equation} \label{eq:app_opt_eq}
\begin{aligned}
\boldsymbol{B}\cdot \hat{G}^{(12)}\boldsymbol{B} &=
g_{1}^{(1)} g_{1}^{(2)} B_x^2+
g_{2}^{(1)} g_{2}^{(2)} B_y^2+
g_{3}^{(1)} g_{3}^{(2)} B_z^2 \\
&=0\,.
\end{aligned}
\end{equation}
This problem is actually equivalent to that of dephasing sweet spots discussed in Ref. \cite{Mauro2024}. The signature of $\hat{G}^{(12)}$, defined as the triplet $(p,n,z)$ of the number of positive ($p$), negative ($n$), and zero ($z$) eigenvalues of this matrix, determines the geometrical nature of the solution(s). There is a series of possible scenarios. If $p=3$ (resp. $n=3$), then $G^{(12)}$ is positive (resp. negative) definite. Therefore, $\boldsymbol{B}\cdot \hat{G}^{(12)}\boldsymbol{B}$ is never zero, and there is no magnetic field orientation where the Larmor vectors of the two dots are orthogonal to each other. This includes the case $r_{\ell}^{(1)}=1/r_{\ell}^{(2)}<1.54$ ($\ell_c^0=16$\,nm) discussed in the main text, where $\alpha$ is maximum in-plane for $\varphi=\pm 45\degree$ but remains $<90\degree$ (see Fig. \ref{fig:sq_ang}c). Note that these maxima move to other $\varphi$'s if the two QDs have different squeezings. 

If $p=1$ (or $n=1$) and $z=2$, a magnetic field pointing anywhere in the plane spanned by the null eigenvectors of $\hat{G}^{(12)}$ achieves the optimal $\alpha=90\degree$. This is indeed the case we target in the main text with $r_{\ell}^{(1)}=1/r_{\ell}^{(2)}=1.54$ ($\ell_c^0=16$\,nm), for which $g_2^0=0$ and $\alpha^0=90\degree$ for any magnetic field in the $xy$ plane. Note, however, that $\boldsymbol{B}$ should not be parallel to $\boldsymbol{x}$ or $\boldsymbol{y}$, as the Larmor frequency of one dot is then zero. Alternatively, if only one of the two QDs is squeezed to reach $g_2^0=0$, then $p=2$ (or $n=2$) and $z=1$, in which case the optimal orientation is $\boldsymbol{B}\parallel\boldsymbol{v}_2^0$, yet not exploitable for the same reason.

The remaining scenario is $p=2$ and $n=1$ (or vice-versa). In this case, the locus of magnetic fields that satisfy $\cos\alpha=0$ forms two cones. The intersection of these cones with the unit sphere $|\boldsymbol{B}|=1$ are two closed curves (lines). This is for example the case when $r_{\ell}^{(1)}=1/r_{\ell}^{(2)}>1.54$ ($\ell_c^0=16$\,nm). We illustrate the existence of these ``optimal lines'' in Fig. \ref{fig:app_optimal2}a, for a pair of QDs with $r_{\ell}^{(1)}=1/r_{\ell}^{(2)}=r_\ell^0=2$. The position of the lines is actually independent of $\varphi$ (as $g_{1}^{(1)}g_{1}^{(2)}=g_{2}^{(1)} g_{2}^{(2)}=g_1^0g_2^0$), and slightly out-of-plane [they appear where $g_1^0g_2^0(B_x^2+B_y^2)=(g_3^0)^2B_z^2$]. The optimal operation points identified in the $\theta$ traces of Fig. \ref{fig:sq_ang}c ($r_\ell^0=2$) are in fact the intersections of the $\varphi=45\degree$ plane with these optimal lines. Fig. \ref{fig:app_optimal2}b highlights the appearance of these lines in orthogonally squeezed QDs with $r_\ell^0>1.54$, and shows how their position $\theta=90\degree\pm\delta\theta_{\rm opt}^0$ depends on $r_\ell^0$ due to the modulation of $g_1^0g_2^0$. The elevation $\delta\theta_{\rm opt}^0$ remains nonetheless very small and only reaches $\delta\theta_{\rm opt}^0\approx 0.4\degree$ for $r_\ell^0=2$. We want to highlight that $\delta\theta_{\rm opt}^0$ is ruled by the $g$ factor anisotropy, and that farther out-of-plane orientations could likely be engineered in QDs with smaller $g_3$, notably in bulk Ge heterostructures \cite{Mauro2025,Costa2025}.

A case that also fulfills $p=2$ and $n=1$ is when only one of the QDs is sufficiently squeezed to switch the sign of a principal in-plane $g$ factor (namely $r_{\ell}^{(1)}< 1.54$, $1/r_{\ell}^{(2)}>1.54$ for $\ell_c^0=16$\,nm). In that case, the optimal lines form closed loops that cross $\theta=90\degree$, as shown in Fig. \ref{fig:app_optimal}a for $r_{\ell}^{(1)}=2$, $r_{\ell}^{(2)}=1$. Interestingly, the angles $\varphi$ of these intersections are also electrically tunable, as they depend on $r_{\ell}^{(1)}$ and $r_{\ell}^{(2)}$ (see Fig. \ref{fig:app_optimal}b). Similar to the recently demonstrated electrical alignment of the sweet spots of neighboring qubits \cite{bassi2024}, this electrical tunability could enable the optimal $\alpha=90\degree$ operation across the QD pairs of a large-scale array.

\section{Effect of confinement on spin manipulation by hopping in circular QDs}\label{app:dep_conf}

\begin{figure}[t!]
\centering
\includegraphics[width=0.98\columnwidth]{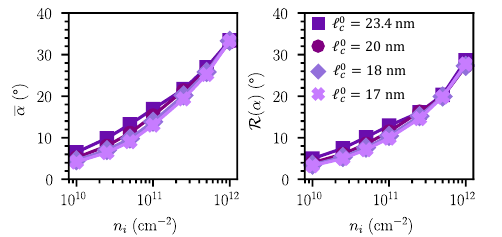}
\caption{Dependence of $\overline{\alpha}$ and $\mathcal{R}(\alpha)$ on the charge trap density $n_i$ for nominally circular QDs with different sizes $\ell_c^0$.}
\label{fig:app_depConf}
\end{figure}

In this Appendix, we discuss the dependence of $\alpha$ on the dot size $\ell_c^0$ for nominally circular QDs. We show that the dot size has little impact on the statistics, contrary to squeezed QDs where confinement plays a more prominent role (see Fig. \ref{fig:sq_ideal}). 

In Fig. \ref{fig:app_depConf} we plot the dependence of $\overline{\alpha}$ and $\mathcal{R}(\alpha)$ on the charge trap density $n_i$ for different dot sizes $\ell_c^0$. We observe that all sizes behave very similarly, especially at large $n_i$. Indeed, for nominally circular QDs the in-plane orientation $\varphi_v$ of the magnetic axes is essentially random regardless of $\ell_c^0$. The dependence of $g_1$ and $g_2$ on the size of the QD makes minor contributions to the variations of $\overline{\alpha}$ at the scale of the strong dependence on $n_i$. Even though there may be specific confinements that maximize $\alpha$ for a given realization of the disorder, there is no significant statistical improvement in working with smaller or bigger QDs.  

\bibliography{var}

\end{document}